\documentclass[journal]{IEEEtran}
\usepackage{cite}
\usepackage{color}
\usepackage{amsmath,amssymb,amsfonts}
\usepackage{algorithmic}
\usepackage{array}
\usepackage{graphicx}
\usepackage{subfigure}
\usepackage{graphicx,times,amsmath,epsfig,graphicx,amsmath} 
\usepackage{subfigure}
\usepackage{caption}
\usepackage{threeparttable}
\usepackage{array,color}
\usepackage{multicol}
\usepackage{multirow}
\usepackage{color}
\usepackage[linesnumbered,boxed,ruled,vlined]{algorithm2e}
\newtheorem{definition}{\textbf{Definition}}

\begin{document}
\title{Perturbed Adaptive Belief Propagation Decoding for High-Density Parity-Check Codes}
\author{Li~Deng,~ Zilong~Liu,~Yong~Liang~Guan,~Xiaobei~Liu,~Chaudhry~Adnan Aslam,~Xiaoxi~Yu,~and Zhiping Shi


\thanks{Li Deng and Zhiping Shi are with the National Key Laboratory on Communications, University of Electronic Science and Technology of China, Chengdu, China; Li Deng is also with School of Electronic Information and Automation, Guilin University of Aerospace Technology, Guilin, China; Zhiping Shi is also with Science and Technology on Communication Networks Laboratory, Shijiazhuang, China
 (E-mail:{\tt dengli@std.uestc.edu.cn; szp@uestc.edu.cn}).}

\thanks{Zilong Liu is with the School of Computer Science and Electrical Engineering, University of Essex, United Kingdom (E-mail: {\tt zilong.liu@essex.ac.uk}).}

\thanks{Yong Liang Guan, Xiaobei Liu, and Xiaoxi Yu are with the
School of Electrical and Electronic Engineering, Nanyang Technological
University, Singapore (E-mail: {\tt \{eylguan,xpliu\}@ntu.edu.sg; XIAOXI001@e.ntu.edu.sg}).}

\thanks{Chaudhry Adnan Aslam is with the public sector R\&D organization, Pakistan (E-mail: {\tt engr\_adnan\_aslam@hotmail.com}).}
}


\maketitle
 \begin{abstract}
Algebraic codes such as BCH code are receiving renewed interest as their short block lengths and low/no error floors make them attractive for ultra-reliable low-latency communications (URLLC) in 5G wireless networks. This paper aims {at} enhancing the traditional adaptive belief propagation (ABP) decoding, which is a soft-in-soft-out (SISO) decoding for high-density parity-check (HDPC) algebraic codes, such as Reed-Solomon (RS) codes, Bose-Chaudhuri-Hocquenghem (BCH) codes, and product codes. The key idea of traditional ABP is to sparsify certain columns of the parity-check matrix corresponding to the least reliable bits with small log-likelihood-ratio (LLR) values. This sparsification strategy may not be optimal when some bits have large LLR magnitudes but wrong signs. Motivated by this observation, we propose a Perturbed ABP (P-ABP) to incorporate a small number of unstable bits with large LLRs  into the sparsification operation of the parity-check matrix.  In addition, we propose to apply partial layered scheduling or hybrid dynamic scheduling to further enhance the performance of P-ABP. Simulation results show that our proposed decoding algorithms lead to improved error correction performances and faster convergence rates than the prior-art ABP variants.
\end{abstract}

\IEEEpeerreviewmaketitle

\section{Introduction}
\IEEEPARstart{R}{eed}-Solomon (RS) codes \cite{RS1960} and Bose-Chaudhuri-Hocquenghem (BCH) codes are high-density parity-check (HDPC) codes with large minimum Hamming distance and no or low error floors.  They have been widely applied in data transmission, broadcasting, and digital storage systems \cite{Wicker1994,Gross2006,xu2007}, etc. Recently, it has been shown that BCH codes outperform short block-length polar codes and low-density parity check (LDPC) codes, with decoding error rates {close to the coding bounds in the finite block-length regime} \cite{Wonterghem2016,URLLC2019}. Hence, BCH codes may be an excellent candidate for the support of ultra-reliable low-latency communications (URLLC) \cite{Durisi2016} featuring short-packet transmissions in 5G networks and beyond.

\subsection{State-of-the-Art of ABP Algorithms}
In most practical systems, algebraic hard-decision decoding (HDD) is adopted for the decoding of RS and BCH codes, such as the syndrome decoding of Berlekamp-Massey algorithm  \cite{Massey1969} and the list decoding of Guruswami-Sudan algorithm \cite{Sudan1999}. Compared with HDD, soft-decision decoding (SDD) is capable of achieving better error correction performance by using the reliability information from the channel  \cite{Mani2019}. Known SDD algorithms for algebraic HDPC codes include the generalized minimum distance  decoding \cite{Forney1966}, Chase decoding \cite{Chase1972}, Chase-Pyndiah decoding for product codes \cite{pyndiah1998}, algebraic soft decoding (ASD)\cite{Kamiya2001,Koetter2003}, ordered statistic decoding \cite{OSD1995, OSD2008}, and adaptive belief propagation (ABP) decoding \cite{Jiang2004,Jiang2006,Baldi2008}, etc. This paper focuses specifically on the ABP for soft-in-soft-out (SISO) decoding.

It is widely recognized that straightforward application of BP decoding to HDPC codes could lead to poor error correction performance, due to a large number of short cycles in the corresponding Tanner graph (TG) \cite{Shayegh2009}. Denote by $N$ and $K$ the codeword length and the message length, respectively. Also, let $M=N-K$ be the number of parity check bits. The parity-check matrix  has dimension of $M \times N$.
To circumvent the limitation of short-cycle overwhelmed BP decoding, the ABP adaptively sparsifies the $M$ parity-check columns associated to $M$ {unreliable bits}, using Gaussian elimination (GE), before performing BP decoding in each iteration. To further improve ABP decoding performance, various improved algorithms have been proposed, such as  hybrid ABP-ASD \cite{Khamy2006}, and stochastic ABP \cite{StochasticABP2008}. A decoding approach similar to ABP is also introduced for BCH codes with simplified parity-check matrix  adaptation \cite{Baldi2008}. In \cite{Jego2009}, a Turbo-oriented ABP called TAB is proposed for product codes, which provides a performance close to that of Chase-Pyndiah algorithm \cite{pyndiah1998}.

\textit{Key Observation}: Among the ABP and its variants, the $M$ bits with smallest log-likelihood ratio (LLR) magnitudes, also called {least reliable bits} in this work, are selected as the $M$ {unreliable bits}. These {unreliable bits}, each connected with only one edge in the corresponding TG (after GE), are highly dependent on the $K$ remaining bits\footnote{In contrast, these $K$ remaining bits have relatively large LLR magnitudes.} to attain improved LLRs. Hence, it is important that these $K$ remaining bits all have correct signs.  Such an unreliable-bits selection strategy is, however, not optimal if some of the remaining bits have wrong signs. In particular, those bits with relatively large LLRs but alternating signs before and after an update in BP decoding tend to be unstable \cite{IVCRBP2011} and hence it is desirable to identify their locations and sparsify their parity-check columns in GE. This key observation motivates us to propose an improved ABP with enhanced unreliable-bits selection strategy.

\textit{Remark}: Throughout this work, we differentiate the following three types of bits:
\begin{itemize}
\item \textit{Unreliable bits}: The $M$ bits whose parity-check columns are to be sparsified before every BP decoding.
\item \textit{Least reliable bits}: The $M$ bits with smallest LLR magnitudes before every BP decoding. In traditional ABP, unreliable bits are the $M$ least reliable bits.
\item \textit{Unstable bits}: Certain bits which have large LLR magnitudes and display alternating signs. 
\end{itemize}

\subsection{Fixed and Dynamic Scheduling for ABP Decoding}
Decoding iteration scheduling is another strategy to improve ABP decoding. Recent advances in the decoding of LDPC codes show that the scheduling strategy has considerable impacts on the decoding performance.  Among a series of major fixed scheduling based BP algorithms, both the layered BP and shuffled BP can attain two times faster convergence rates and comparable error correction performances compared to the flooding BP \cite{Sharon2004,Juntan2002}. {The authors of \cite{Aslam2017} presented} an edge-based flooding schedule (called ``e-Flooding") for RS codes by only partially updating the edges originating from the less reliable check nodes for complexity reduction. 

Compared with the fixed scheduling, dynamic scheduling is able to give further improvement to BP decoding, if run-time processing can be afforded\cite{Casado2010,IVCRBP2011,SVNF2013,rel_RBP2015,OVRBP2015,DSVNF2017}. The residual based dynamic schedules, which update the edge messages with the maximum residual first, are capable of circumventing performance deterioration incurred by trapping sets \cite{trapping_sets}. A layered residual BP (LRBP) is introduced in \cite{Lee2015} for ABP based decoding of RS codes, which updates the unreliable bits more frequently in each iteration with a sequential updating order of check nodes (CNs) to achieve better error correction performance. {The authors of \cite{Lee2015} further presented a double-polling residual BP (DP-RBP) \cite{Lee2017} by selecting the variable nodes (VNs) and CNs with a double-polling mechanism, where the least reliable VNs and high-degree CNs have more opportunities to get updated. However, both of the LRBP and DP-RBP decoders could not be able to prevent the silent-variable-nodes which have no chance to get updated during the decoding \cite{SVNF2013}}.


In view of the above background, it is instructive to explore scheduling strategies for ABP. However, we found that straightforward applications of the existing fixed or dynamic scheduling may not be effective as they are not optimized for HDPC codes.

\subsection{Contributions}
This work aims for enhancing the traditional ABP to achieve better error correction performances and faster decoding convergence rates for HDPC codes. Our main novelty and contributions are summarized as follows:
\begin{enumerate}
\item We propose a Perturbed ABP (P-ABP) algorithm which constitutes a refined {unstable-bits} selection strategy. We select a small number of bits (denoted by $\rho$) with relatively large LLRs, in addition to the $(M-\rho)$ {least reliable bits}, to be included in the parity-check matrix sparsification. We develop a mechanism to first identify those {unstable bits} displaying relatively large LLRs but alternating signs in the BP iterations, and include them as part of the $\rho$ bits. We further present an extrinsic  information  analysis method to obtain {good values of $\rho$}.\footnote{Note that the traditional ABP decoding may be regarded as a special case of the proposed P-ABP decoding with $\rho=0$. A good $\rho$ leads to better decoding error probability as well as faster convergence rate.}
\item We develop a partial layered  scheduling for P-ABP (called PL-P-ABP) which can well adapt to the systematic structure of HDPC matrix after GE. The proposed partial layered message updating strategy can also skip certain edges associated with short cycles to stop unreliable message passing, leading to the significantly improved convergence rate and comparable or better error correction performance than the flooding and shuffled scheduling schemes. We also apply a variation of PL-P-ABP to decode product codes.
\item For decoders that can afford more run-time computations, we present a hybrid dynamic scheduling for P-ABP (called as HD-P-ABP) aiming  for faster convergence rate than the proposed PL-P-ABP. HD-P-ABP combines the merits of layered scheduling and dynamic silent-variable-node-free (D-SVNF) scheduling to avoid multiple types of greedy groups\footnote{A greedy group refers to a few nodes in the TG of a code excessively consuming computation resources in BP decoding. Such a greedy group is a barrier to optimal decoding as the beliefs of certain other nodes may never get updated.} occurring in traditional dynamic scheduling based decoding algorithms. 
\end{enumerate}

\subsection{Organization}
The remainder of this paper is organized as follows. Section II provides some preliminaries of BCH/RS codes, BP decoding with different fixed scheduling strategies, and the rationale of ABP. Section III describes the proposed P-ABP, together with the proposed partial layered scheduling and hybrid dynamic scheduling for P-ABP. Section IV gives the complexity analysis of the proposed decoding algorithms. The simulation results and relevant discussions are presented in Section V, followed by conclusions in Section VI. {The list of acronyms used in this paper is shown in Table \ref{table_0}}.
\begin{table}[!t]
\centering
\newcommand{\tabincell}[2]{\begin{tabular}{@{}#1@{}}#2\end{tabular}}
\caption{List of acronyms}
\label{table_0}
\setlength{\tabcolsep}{3pt}
\begin{tabular}{|p{50pt}|p{150pt}|}
\hline
Acronyms  & Descriptions \\
\hline

ABP       & adaptive belief propagation \\
ASD       & algebraic soft decoding\\
AWGN      & additive white Gaussian noise           \\
BCH       & Bose-Chaudhuri-Hocquenghem \\
BER       & bit error rate\\
BP        & belief propagation \\
BPSK      & binary-phase-shift-keying\\
CNs       & check nodes\\
D-SVNF    & dynamic silent-variable-node-free scheduling\\
e-Flooding& edge-based flooding scheduling\\
EXIT      & extrinsic information transfer\\
FER       & frame error rate\\
GE        & Gaussian elimination \\
HDPC      & high-density parity-check    \\
HDD       & hard-decision decoding \\
HD-P-ABP  & P-ABP with hybrid dynamic scheduling \\
LDPC      & low-density parity-check\\
LLR       & log-likelihood-ratio\\
LRBP      & layered residual BP \\
ML        & maximum likelihood \\
P-ABP     & Perturbed ABP \\
PL-P-ABP  & P-ABP with partial  layered  scheduling \\
PL-P-ABP-P& PL-P-ABP for product codes\\
PUM       & perturbed unreliable-bits mapping \\
RS        & Reed-Solomon\\
SISO      & soft-in-soft-out\\
SDD       & soft decision decoding\\
SNR       & signal-to-noise-ratio\\
TAB       & Turbo-oriented ABP \\
TG        & Tanner graph \\
URLLC     & ultra-reliable low-latency communications  \\
VNs       & variable nodes\\
\hline
\end{tabular}
\end{table}
\section{Preliminaries}
Throughout this paper, denote by $N$ and $K$ (as given in Section I) the codeword length and the message length of a code, respectively. Also, let $M=N-K$ be the number of parity-check bits.
\subsection{{Bose-Chaudhuri-Hocquenghem (BCH) and Reed-Solomn (RS) Code Family}}
\subsubsection{BCH codes}
{The BCH code forms a large class of powerful linear cyclic block codes, which is widely known for its capability of correcting multiple errors and simple encoding/decoding mechanisms. The BCH code over the base field  of GF ($q$) can be defined by a parity-check matrix  $\mathbf{H}$ over the extension field of GF ($q^m$) which is shown below:}  
\begin{equation}
{\bf{H}} = \left[ {\begin{array}{*{20}{c}}
1&\beta^b &\beta^{2b} & \cdots &{{\beta ^{(N - 1)b}}}\\
1&{{\beta ^{(b+1)}}}  &\beta^{2(b+1)} & \cdots &{{\beta ^{(N - 1)(b+1)}}}\\
 \vdots & \vdots & \ddots & \vdots \\
1&{{\beta ^{b+d-2}}}&{{\beta ^{2(b+d-2)}}}& \cdots &{{\beta ^{(N - 1)(b+d-2)}}}
\end{array}} \right],
\end{equation}
where $\beta$ is an element of GF $(q^m)$ of order $N$, $b$  any integer ($0 \leq b \leq  N$ is sufficient), and $d$ an integer with $2 \leq d  \leq N$. If $\beta$ is a primitive element of GF $(q^m)$, then the codeword length is $N=q^m-1$, which is the maximum possible codeword length for the extension field GF $(q^m)$. 
The parity-check matrix  for a $t$-error-correcting primitive narrow-sense BCH code can be described as
\begin{equation}
{{\bf{H}}_{p}} = \left[ {\begin{array}{*{20}{c}}
1&\beta &\beta^2 & \cdots &{{\beta ^{N - 1}}}\\
1&{{\beta ^2}} &\beta^{4} & \cdots &{{\beta ^{2(N - 1)}}}\\
 \vdots & \vdots & \vdots & \ddots & \vdots \\
1&{{\beta ^{2t}}} &\beta^{4t} & \cdots &{{\beta ^{(2t)(N - 1)}}}
\end{array}} \right].
\end{equation}

In this paper, we consider binary BCH code, where the channel alphabets are binary elements and the elements of the parity check matrix are in GF $(2^m)$ ($q=2$). For any integer $m \geq 3$ and $t < 2^{m-1}$, a binary BCH code can be found with the  length of $N=2^m-1$ and the minimum distance $d\geq 2t + 1$, where $t$ is the error correction power.

\vspace{0.1in}

\subsubsection{RS codes}
The RS code is a BCH code with non-binary elements, where {the base field GF $(q)$ is the same as the extension field GF $(q^m)$, i.e., $m=1$}. 
The generator polynomial of a $t$-error-correcting RS code is
\begin{equation}
g(x)=(x-\beta^b)(x-\beta^{(b+1)})\cdots(x-\beta^{(b+2t-1)}),
\end{equation}
where the minimum distance $d\geq 2t+1$, which is independent of  $\beta$ and $b$. Usually $\beta$ is chosen to be primitive in order to maximize the block length.
The base exponent $b$ can be chosen to reduce the encoding and decoding complexity.

Assuming that the coded sequence $\mathbf{c}$ is passed to the additive white Gaussian noise (AWGN) channel after the binary-phase-shift-keying (BPSK) modulation, the received signal can be given by ${\mathbf{y}} = {\mathbf{x}} + {\mathbf{w}}$,
where $\mathbf{x}$ is the modulated signal of the coded sequence $\mathbf{c}$, and $\mathbf{w}$ is the noise vector with the variance of ${\sigma^2}$. The BP algorithm uses the channel LLR as its input, which can be expressed as
\begin{equation}
\label{Eq_Lch}
{L_\text{ch}}({v_n}) = \log \frac{{p({\bf{y}}\left| {{v_n} = 1)} \right.}}{{p({\bf{y}}\left| {{v_n} = 0)} \right.}} = \frac{2}{{\sigma^2}}{y_n},
\end{equation}
where $y_n$ is the $n$-th element of the received signal $\mathbf{y}$, {and $v_n$ is the $n$-th coded bit of the sequence $\mathbf{c}$, $n=1, 2, \ldots, N$}.

\vspace{0.1in}

\subsection{BP Decoding with Different Scheduling Strategies}
 BP is a classical message passing algorithm which recursively exchanges the belief information between the VNs and CNs in a TG. Before BP is executed, the soft information of each VN is initialized by the channel LLR given in (\ref{Eq_Lch}). BP iteratively propagates the belief information along the edges of TG from CNs to VNs (denoted by $C_{c_m \rightarrow v_n}$, called a C2V message), and from VNs to CNs (denoted by $V_{v_n \rightarrow c_m}$, called a V2C message). For the typical message passing schemes of flooding, shuffled, and layered scheduling, the updating orders of $C_{c_m \rightarrow v_n}$ and $V_{v_n \rightarrow c_m}$ may lead to different decoding performances.

We assume that the binary parity-check matrix  $\mathbf{H}_b$ has a dimension of $M\times N$, the set of VNs that are connected to the $m$-th CN is ${\cal{N}}\left( c_m \right)$, and the set of CNs that are connected to the $n$-th VN is ${\cal {M}} \left(v_n\right)$. ${\cal {N}}\left( c_m \right)\backslash v_n$ denotes the subset of ${\cal {N}}\left( c_m \right)$ without $v_n$ and ${\cal {M}}\left( v_n \right)\backslash c_m$ denotes the subset ${\cal {M}} \left(v_n\right)$ without $c_m$. The three types of fixed scheduling based BP algorithms are described as follows.

\vspace{0.1in}

\subsubsection{Flooding BP algorithm}
\begin{itemize}
  \item Phase 1: C2V message update.

  For $m=1: M$, generate and propagate $C_{c_m \rightarrow v_n}$.
\begin{small}
\begin{equation}\label{Eq:flooding_c2v}
{C^{(i)}_{c_m \rightarrow v_n} = 2 {\tanh}^{-1} \left( \prod\limits_{\substack{ v_{j}\in \mathcal{N}(c_m)\setminus v_n}} {\tanh}\left(\frac{V^{(i-1)}_{v_{j} \rightarrow c_m}}{2}\right) \right)},
\end{equation}
\end{small}

where the superscripts $i$ and $(i-1)$ denote the $i$-th and the $(i-1)$-th decoding iterations, respectively.
\item Phase 2: V2C message update.

  For $n=1: N$, generate and propagate $V_{v_n \rightarrow c_m}$ .
\begin{small}
\begin{equation}\label{Eq:flooding_v2c}
{V^{(i)}_{v_n \rightarrow c_m} = L_\text{ch}(v_n) + \sum\limits_{c_{j}\in \mathcal{M}(v_n)\setminus c_m} C^{(i)}_{c_{j} \rightarrow v_n}}.
\end{equation}
\end{small}

Under the flooding scheduling, the C2V messages update simultaneously at the first half iteration, while all the V2C messages update at the second half.
These steps are iterated until the maximum iteration number is reached or the parity-check (syndrome) equations below are satisfied:
\begin{equation}
\label{eqn_paritycheck}
s_{c_m}^{(i)} = \sum_{n:v_n\in \mathcal{N}(m)} h_{m,n}\ \hat{c}_n^{(i)} = 0,
\end{equation}

where $h_{m,n} \in \mathbf{H}_b^i, 1 \leq m \leq M, 1 \leq n \leq N$, $\hat{c}_n^{(i)} \in \mathbf{\hat{c}}^{(i)}, 1 \leq n \leq N$, and $\mathbf{\hat{c}}^{(i)}$ is the estimated codeword in the $i$-th iteration.
\end{itemize}

\vspace{0.1in}

\subsubsection{Shuffled BP algorithm}
\begin{itemize}
  \item For $n=1: N$, and for each $c_m \in {\cal {M}} \left(v_n\right)$, update the C2V and V2C messages serially for each VN. The V2C update follows (\ref{Eq:flooding_v2c}), while the C2V update can be split into two parts as:
\begin{small}
\begin{equation}
\label{Eq:_shuffled c2v}
  \begin{split}
    C^{(i)}_{c_m \rightarrow v_n} = 2\text{ tanh}^{-1} \left( \prod\limits_{\substack{ v_{j}\in \mathcal{N}(c_m) \\ v_{j}<v_n}} \text{tanh}\left(\frac{V^{(i)}_{v_{j} \rightarrow c_m}}{2}\right) \cdot \right. \\
     \left. \prod\limits_{\substack{ v_{j}\in \mathcal{N}(c_m) \\ v_{j}>v_n}} \text{tanh}\left(\frac{V^{(i-1)}_{v_{j} \rightarrow c_m}}{2}\right)  \right),
  \end{split}
\end{equation}
\end{small}where the first and the second parts represent the V2C messages in the current $i$-th iteration and the previous $(i-1)$-th iteration, respectively.
\end{itemize}

\vspace{0.1in}

\subsubsection{Layered BP algorithm}
\begin{itemize}
  \item For $m=1: M$, and for each $v_n \in {\cal {N}} \left(c_m\right)$, update the C2V and V2C messages serially for each CN.  The C2V update follows (\ref{Eq:flooding_c2v}), while the V2C update can be described as
\begin{small}
\begin{equation}
\label{Eq:_layered v2c}
{
V^{(i)}_{v_n \rightarrow c_m}\!=\! L_\text{ch}(v_n)\! +\! \sum\limits_{\substack{ c_{j}\in \mathcal{M}(v_n) \\ c_{j}<c_m}} C^{(i)}_{c_{j} \rightarrow v_n}
  \!+\! \sum\limits_{\substack{ c_{j}\in \mathcal{M}(v_n) \\ c_{j}>c_m}} C^{(i-1)}_{c_{j} \rightarrow v_n},
  }
\end{equation}
\end{small}

where the last two terms represent the C2V messages in the $i$-th and $(i-1)$-th iterations, respectively. It can seen from (\ref{Eq:_shuffled c2v}) and (\ref{Eq:_layered v2c}) that the shuffled and layered schedules allow a quick access to the latest updated message $V^{(i)}_{v_{j} \rightarrow c_m}$ for $v_{j}<v_n$  and $C^{(i)}_{c_{j} \rightarrow v_n}$ for $c_{j}<c_m$, respectively. That explains why shuffled and layered BP can achieve faster convergence rate than flooding BP for most LDPC codes.
\end{itemize}

\vspace{0.1in}

\subsection{ Traditional ABP Algorithm}
The main idea of traditional ABP decoder is to adaptively sparsify the parity-check columns corresponding to the $M$ {unreliable bits} (which are {least reliable bits} ordered by their LLR magnitudes) with the aid of GE, followed by BP decoding in each iteration. 

Let the LLR of the $n$-th coded bit $v_n$ at the $i$-th iteration be {$L^{(i)}\left( {{v_n}} \right)$, where $n=1,2,\cdots,N$}. Formally, the LLR vector consisting of the $N$ bits can be described as
\begin{equation}
{
{{\bf{L}}^{(i)}} =\left[ {{L^{(i)}}\left( {{v_1}} \right),{L^{(i)}}\left( {{v_2}} \right), \cdots, {L^{(i)}}\left( {{v_N}} \right)} \right].}
\end{equation}

Before the BP decoding starts, ${\bf{L}}^{(0)}$ is initialized by ${{\bf{L}}_\text{ch}}$ from the channel, where ${{\bf{L}}_\text{ch}} = \left[ {{L_\text{ch}}\left( {{v_1}} \right),{L_\text{ch}}\left( {{v_2}} \right), \cdots, {L_\text{ch}}\left( {{v_{N}}} \right)} \right]$ (see (\ref{Eq_Lch})). In each iteration, the ABP algorithm mainly consists of two stages: the parity-check matrix  updating stage and LLR updating stage.

In the parity-check matrix  updating stage, the magnitudes of ${\bf{L}}^{(i)}(v_n)$ are sorted in an ascending order. The $M$ {least reliable bits} are selected and their indices are recorded. Then GE is applied to generate an updated parity-check matrix  ${\bf{\bar {H}}}_b^{(i)}$, where the {least reliable bits} are mapped to the sparse part of ${\bf{ \bar {H}}}_b^{(i)}$ such that every least reliable bit is connected with one CN. The idea is to ``box" and freeze the propagations of belief messages associated to those least reliable bits from affecting the remaining $K$ ones with relatively large LLR magnitudes.

In the LLR updating stage, the flooding BP is adopted to generate the extrinsic LLR associated with ${\bf{\bar {H}}}_b^{(i)}$ as \cite{Jiang2006}
\begin{small}
\begin{equation}\label{Eq:LLR_ext}
L_{\text{ext}}^{(i)}({v_n})\! = \! \sum\limits_{c_m \in {\cal{M}}\left(v_n \right)}\!{2{{\tanh }^{ - 1}}\left( {\prod\limits_{v_j  \in {\cal{N}}\left( c_m \right)\backslash v_n} {\tanh \left( {\frac{{{L^{(i)}}({v_j })}}{2}} \right)} } \right)}.
\end{equation}
\end{small}

The LLR of each bit is then updated by \cite{Jiang2006}
\begin{equation}
{L^{(i+1)}}\left( {{v_n}} \right) = {L^{(i)}}\left( {{v_n}} \right) + \alpha L_{\text{ext}}^{(i)}\left( {{v_n}} \right) \label{graDes},
\end{equation}
where $\alpha  \in (0,1]$ is the damping factor. ${{\bf{L}}^{(i + 1)}}$ will be sent from the VNs to the CNs for the next iteration.
\section{Proposed Perturbed ABP}
In this section, we propose Perturbed ABP (P-ABP) as an enhancement over the traditional ABP. P-ABP includes (i) a perturbed unreliable-bits mapping (PUM) scheme, and (ii) a partial layered scheduling, or a hybrid dynamic scheduling.
\subsection{Proposed PUM}
\subsubsection{Definition of PUM}
The traditional ABP is designed such that every least reliable bit is connected with one CN only. Such a feature helps to ``freeze" the flow of the weak belief message of a least reliable bit from propagating to any other nodes in TG during the BP decoding. 
However, ABP would not be optimal if there exist some unstable bits which have large LLRs but incorrect signs. In particular, bits displaying large LLRs but {alternating signs after every other iteration} tend to be unstable and hence should also be frozen in message passing.  Formally, the definition of PUM is given as follows:
\vspace{0.1in}
\begin{definition}
{PUM} refers to an operation which sparsifies the parity-check  columns corresponding to the first $(M-\rho)$ least reliable bits and $\rho$ number of carefully selected bits, especially to the unstable ones with large LLR magnitudes and alternating signs, where $\rho$ is called the perturbation factor of P-ABP.

\end{definition}

\subsubsection{Description of PUM}
Detailed algorithm of PUM is shown in \textbf{SubAlg-1} with some definitions of the bit index sets given first. Denote by $\varepsilon_j^{(i)}$ the bit index of the $j$-th lowest absolute value in ${\bf{L}}^{(i)}$. The least reliable bit index set (ordered by their LLR magnitudes) at the $i$-th iteration is defined as
\begin{equation}
\mathbf{URL}^{(i)} = \left\{ {{\varepsilon _{1}^{(i)}},{\varepsilon _{2}^{(i)}},\cdots,{\varepsilon _{M}^{(i)}}} \right\},
\end{equation}
where its complementary bit index set is
\begin{equation}
\textbf{RL}^{(i)} =\{1,2,\cdots,N\}\setminus \mathbf{URL}^{(i)}= \left\{ {{\varepsilon _{M + 1}^{(i)}},{\varepsilon _{M + 2}^{(i)}},\cdots,{\varepsilon _{N}^{(i)}}} \right\}.
\end{equation}
$\mathbf{URL}^{(i)}_{1}$ and $\mathbf{URL}^{(i)}_{2}$ are index subsets of the first $M-\rho$  bits and the last $\rho$ bits in $\mathbf{URL}^{(i)}$, i.e.,
\begin{equation}
\mathbf{URL}^{(i)}_{1} = \left\{ {{\varepsilon _{1}^{(i)}},{\varepsilon _{2}^{(i)}},\cdots,{\varepsilon _{M-\rho}^{(i)}}} \right\},
\end{equation}


\begin{equation}
\mathbf{URL}^{(i)}_{2}  = \left\{ {{\varepsilon _{M-\rho+1}^{(i)}},{\varepsilon _{M-\rho+2}^{(i)}},\cdots,{\varepsilon _{M}^{(i)}}} \right\}.
\end{equation}

\begin{figure}[!t]
  \centering
  \includegraphics[width=80mm,height=80mm]{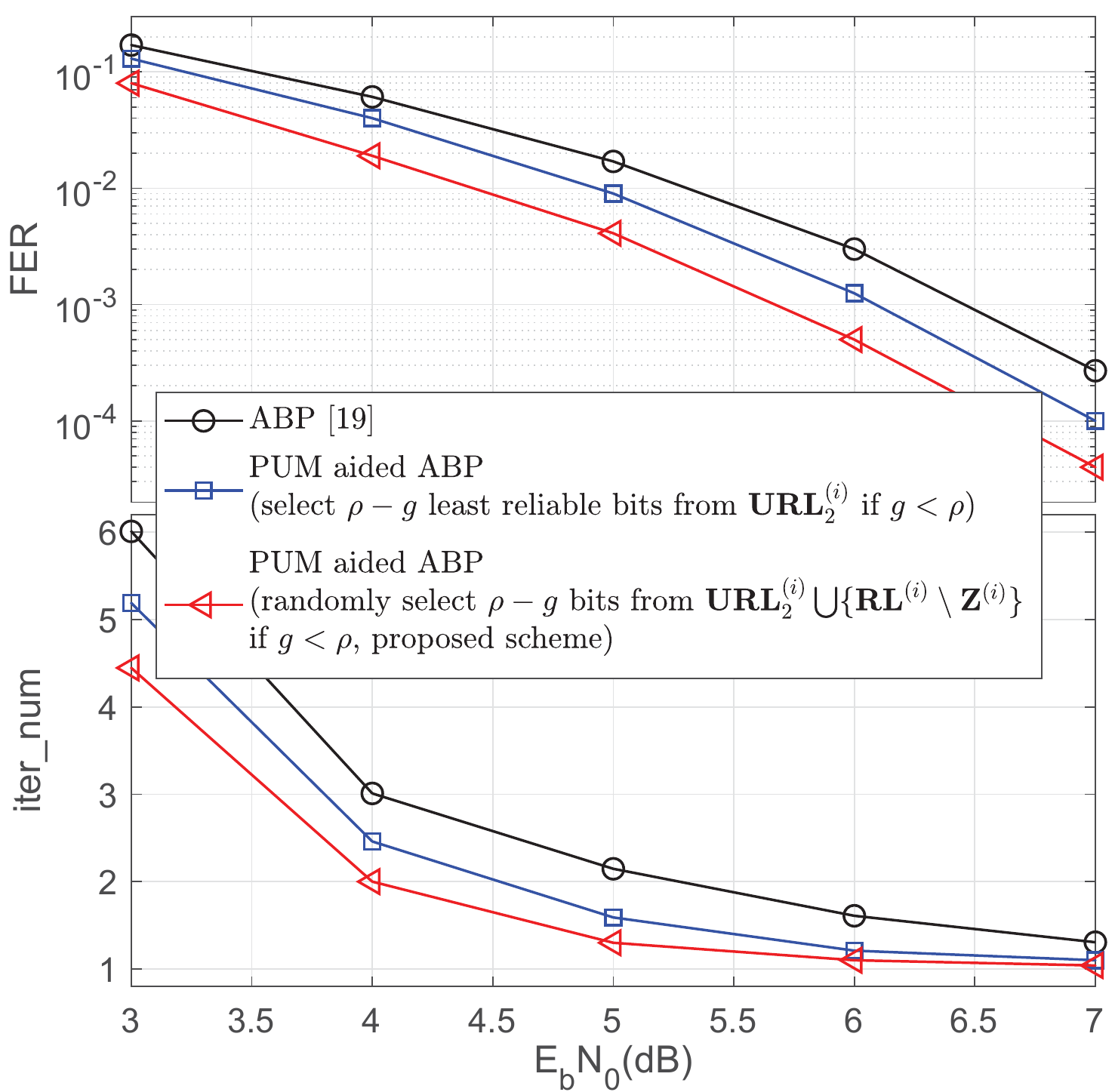}
   \caption {{Comparison of PUM with different selection strategies when $g<\rho$ for (127, 92)-BCH code with $i_{\max}=20$.}}
   \label{fig:PUM comp}
\end{figure}

For a given perturbation factor $\rho$, i.e., the number of perturbed bits,  the main task of \textbf{SubAlg-1} is to generate a refined bit index set $\widehat{\mathbf{URL}}^{(i)}$ at the $i$-th iteration which specifies the $M$ selected parity-check columns for sparsification with the aid of GE. The refined $\widehat{\mathbf{URL}}^{(i)}$ can be generated based on the LLRs of the latest two iterations (i.e., ${{\bf{L}}^{(i)}}$ and ${{\bf{L}}^{(i-1)}}$). The alternating-sign bits (from the $K$  bits with largest LLR magnitudes) are first detected, then their corresponding parity-check columns are sparsified. {If the total number of alternating-sign bits $g$ is less than $\rho$ in the $i$-th iteration, then the least reliable bits in $\mathbf{URL}_2^{(i)}$  and some other bits in $\{\mathbf{RL}^{(i)}\setminus\mathbf{Z}^{(i)}\}$ with large LLR magnitudes are randomly selected for sparsification of their parity-check columns.} {The reason of random selection from $\mathbf{URL}_{2}^{(i)} \bigcup \{\mathbf{RL}^{(i)}\setminus\mathbf{Z}^{(i)}\}$ rather than directly selecting the least reliable bits in $\mathbf{URL}_{2}^{(i)}$ lies in that the former may help bring in more LLR diversity into the P-ABP decoder. 
 This is verified in Fig. \ref{fig:PUM comp} which shows the comparison of PUM aided ABP with random selection versus selecting lowest-LLR bits in $\mathbf{URL}_2^{(i)}$. The latter refers to the scheme that when $g<\rho$, $\rho-g$ least reliable bits in $\mathbf{URL}_2^{(i)}$ with lower LLR magnitudes are selected into the refined $\widehat{\mathbf{URL}}_{2}^{(i)}$. The compared algorithms stop to work when a successful decoding is attained within the maximum iteration number ($i_{\max}$), and the average decoding iteration number $iter\_{num}$ is used to estimate the decoding convergence rate. 
As shown in Fig. \ref{fig:PUM comp}, the PUM aided ABP with random selection (our proposed scheme) outperforms the original ABP and the compared scheme in both the frame error rate (FER) and decoding convergence.} An approach for the finding of a good value of $\rho$ will be presented later.

 \begin{figure*}[htbp]
  \centering
    \includegraphics[width=185mm,height=60mm]{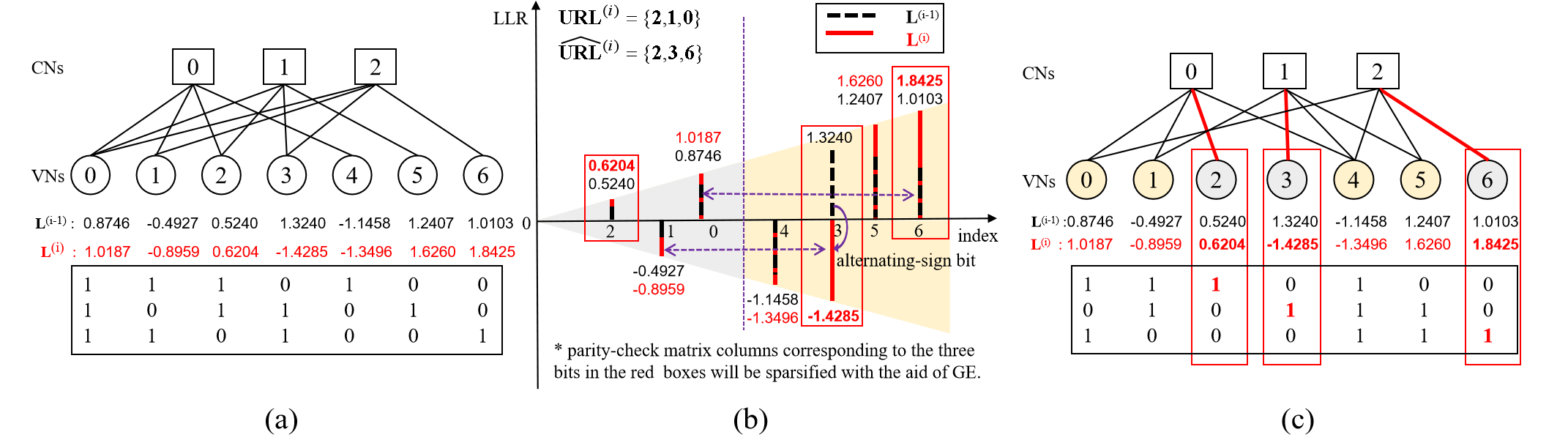}
   \caption {Illustration of the proposed PUM for a (7, 4)-Hamming code and $\rho=2$: (a) TG and parity-check matrix  before GE; {(b) How PUM is carried out;} (c) TG and parity-check matrix  after GE.}\label{fig:Perturbation}
\end{figure*}

\begin{figure*}[!t]
  \centering
    \includegraphics[width=170mm,height=40mm]{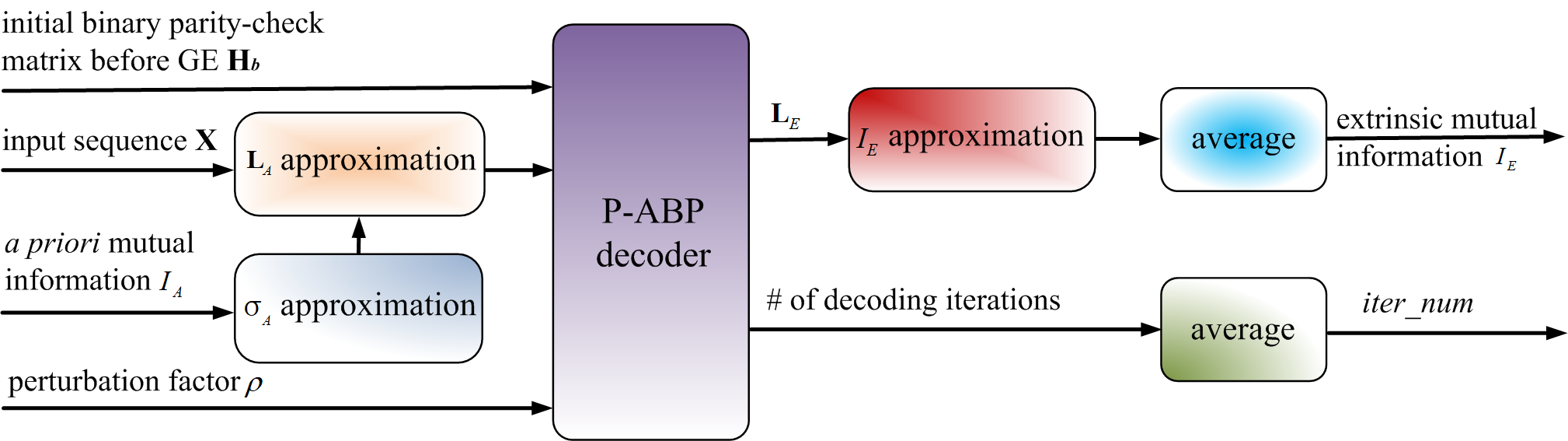}
   \caption {{Extrinsic information analysis for perturbation factor $\rho$ of the proposed P-ABP decoder.}}
   \label{fig:EXIT_P}
\end{figure*}

\begin{algorithm}[!ht]
\NoCaptionOfAlgo
\caption{\textbf{SubAlg-1}: Proposed PUM Algorithm}
\KwIn{$\mathbf{L}^{(i)}$, $\mathbf{L}^{(i-1)}$;}
\KwOut{Refined bit index set $\widehat{\mathbf{URL}}^{(i)}$ for sparsification of their parity-check columns.}

Step 1: Select $\rho$ using off-line extrinsic information analysis.

Step 2: Record the LLR signs and absolute values before and after each decoding iteration;

Step 3: Identify those bits from $\textbf{RL}^{(i)}$ with LLR sign changes and arrange their indices in descending order of amplitudes in $\mathbf{Z}^{(i)}=\{z_1^{(i)},z_2^{(i)},\cdots,z_g^{(i)}\ | \ g \leq K \}$;

Step 4: Select $\rho$ bits and denote their index set by $\widehat{\mathbf{URL}}_{2}^{(i)}$:

  \If {$g \geq \rho$}
    {choose the first $\rho$ entries from $\mathbf{Z}^{(i)}$ to generate $\widehat{\mathbf{URL}}_{2}^{(i)}$;}

  \ElseIf {$0< g < \rho$}
  {{choose all the alternating-sign bits from $\mathbf{Z}^{(i)}$, and then randomly choose $\rho-g$ bits from $\mathbf{URL}_{2}^{(i)} \bigcup \{\mathbf{RL}^{(i)}\setminus\mathbf{Z}^{(i)}\}$ to generate $\widehat{\mathbf{URL}}_{2}^{(i)}$;}}

  \Else 
  {randomly choose $\rho$ bits from $\mathbf{URL}_{2}^{(i)} \bigcup \mathbf{RL}^{(i)}$ to generate $\widehat{\mathbf{URL}}_{2}^{(i)}$;}

Step 5: Generate the new refined bit index set as $\widehat{\mathbf{URL}}^{(i)}=\mathbf{URL}_{1}^{(i)} \bigcup \widehat{\mathbf{URL}}_{2}^{(i)}$.
\end{algorithm}


Fig. \ref{fig:Perturbation} illustrates the PUM process by employing a (7, 4)-Hamming code as a simple example, where $N=7$, $M=3$, and the perturbation factor is assumed to be $\rho=2$. Fig.  \ref{fig:Perturbation} (a) shows its TG, parity-check matrix  and LLRs at the $i$-th iteration $\mathbf{L}^{(i)}$ and the $(i-1)$-th iteration $\mathbf{L}^{(i-1)}$.  Fig. \ref{fig:Perturbation} (b) describes how  the perturbation is carried out.   Firstly, ${\bf{L}}^{(i)}$ are sorted in an ascending order according to the  amplitudes. In this example, $\mathbf{URL}^{(i)} = \left\{2,1,0 \right\}$, $\mathbf{URL}^{(i)}_{1} = \left\{2 \right\}$, $\mathbf{URL}^{(i)}_{2} = \left\{1, 0 \right\}$, and $\mathbf{RL}^{(i)} = \left\{4,3,5,6\right\}$. Secondly, we detect the bits with alternating signs in $\mathbf{RL}^{(i)}$ based on $\mathbf{L}^{(i)}$ and $\mathbf{L}^{(i-1)}$, and then
put their indices in set $\mathbf{Z}^{(i)}$. In  Fig. \ref{fig:Perturbation} (b), only bit $v_3$ has LLRs with reversed signs during the latest two iterations, thus $\mathbf{Z}^{(i)}=\left\{3\right\}$. {Since the number of bits with alternating signs is one, which is smaller than $\rho$, then one bit, say $v_6$, is randomly selected from $\mathbf{URL}_{2}^{(i)} \bigcup  \{\mathbf{RL}^{(i)}\setminus\mathbf{Z}^{(i)}\}$. Therefore, we have $\widehat{\mathbf{URL}}_{2}^{(i)}=\left\{3,6\right\}$. Finally, the refined bit index set whose corresponding parity-check  columns will be sparsified can be written as $\widehat{\mathbf{URL}}^{(i)}=\mathbf{URL}^{(i)}_{1} \bigcup \widehat{\mathbf{URL}}^{(i)}_{2}=\left\{2,3,6\right\}$. } In Fig. \ref{fig:Perturbation} (c), GE is implemented, where the bits in $\widehat  {\mathbf{URL}}^{(i)}$ are mapped to the sparse part of $\mathbf{\bar{H}}_b^{(i)}$.

\subsubsection{Perturbation Factor $\rho$}
In this subsection, we propose an extrinsic information analysis method to find a good perturbation factor $\rho$. {For a given initial binary
parity-check matrix $\mathbf{H}_b$ and \textit{a priori} mutual information $I_A$,
the purpose here is to find a good $\rho$ which leads to the maximum extrinsic
information $I_E$ as well as the smallest average iteration number $iter\_{num}$. Note that the conventional EXIT chart analysis \cite{Brink2001} generally assumes a fixed Tanner graph. In our proposed P-ABP decoder, however, the Tanner graph changes according to the obtained LLRs over different BP iterations. In order to conduct the extrinsic information analysis, we treat the whole P-ABP decoder as a black box (as shown in Fig. \ref{fig:EXIT_P}), and do not care how the extrinsic information is transferred among different nodes in the intermediate parity-check matrices during the BP iterations. We only evaluate the resultant extrinsic information $I_E$ and $iter\_{num}$ after the decoding process. Finally, the $\rho$ with the maximum $I_E$ and the smallest $iter\_{num}$ is selected for a certain $I_A$.}

Fig. \ref{fig:EXIT_P} shows the signal processing flow of extrinsic information analysis. For a given input sequence $\mathbf{X}$ and $I_A$, \textit{a priori} LLR $\mathbf{L}_A$ is generated to feed into the P-ABP decoder with a certain $\rho$ for extrinsic information analysis. First, the standard deviation $\sigma_A$ of $I_A$ is approximated as \cite{Brannstrom2005}

\begin{equation} \label{Eq:Ia2sigma}
 \sigma_A= J^{-1}(I_A) \approx  \left(-\frac{1}{H_1} \log_2\left (1-I_A^{\frac{1}{H_3}}\right )\right)^{\frac{1}{2H_2}},
\end{equation}
where $H_1=0.3073$, $H_2=0.8935$, and $H_3=1.1064$. Then, the \textit{a priori} LLR $\mathbf{L}_A$ can be estimated with the approximation method as \cite{Hanzo2002}

\begin{equation} \label{Eq:L_A}
 \mathbf{L}_A=\frac{\sigma^2_A}{2} (2\mathbf{X}-1)+\sigma_A \mathbf{L}_0 ,
\end{equation}
where {$\sigma^2_A$ is the variance of $I_A$}, and $\mathbf{L}_0$ denotes the noise term, which is a sequence with the initial distribution of $\mathcal{N}(0, 1)$. However, the noise term is expected to have the standard deviation of $\sigma_A$, thus $\mathbf{L}_0$ is multiplied by $\sigma_A$.

Finally, $I_E$ is approximated by the {extrinsic} LLR ($\mathbf{L}_E$) of the proposed P-ABP decoder \cite{Hagenauer2004}
\begin{equation} \label{Eq:LLR2IM}
\begin{split}
 I_E(L;X)=& 1-\mathbf{E}\left \{\log_2(1+e^{-\mathbf{L}_E}) \right \}\\
      \approx & 1-\frac{1}{N} \sum_{n=1}^{N}\log_2\left (1+e^{x_nl_n} \right ) \\
      \approx & 1-\frac{1}{N} \sum_{n=1}^{N}\tilde{H}_b(P_{e_n}),
 \end{split}
\end{equation}
where $l_n$ denotes the $n$-th element of $\mathbf{L}_E$, i.e., the LLR value of the $n$-th bit $x_n$, $P_{e_n}$ is the probability of $x_n \cdot \text{sgn}(l_n)=-1$, i.e.,
\begin{equation*}
 P_{e_n}=\frac{e^{+\frac{|l_n|}{2}}}{e^{+\frac{|l_n|}{2}}+e^{-\frac{|l_n|}{2}}},
\end{equation*}
 and $\tilde{H}_b$ denotes the binary entropy function:
\begin{equation*}
 \tilde{H}_b(P_{e_n})=-P_{e_n}\cdot\log_2(P_{e_n})-(1-P_{e_n})\cdot\log_2(1-P_{e_n}).
\end{equation*}

\begin{figure}[!t]
  \centering
   \subfigure[]{
    \label{fig:selected P_IA}
     \includegraphics[width=80mm,height=65mm]{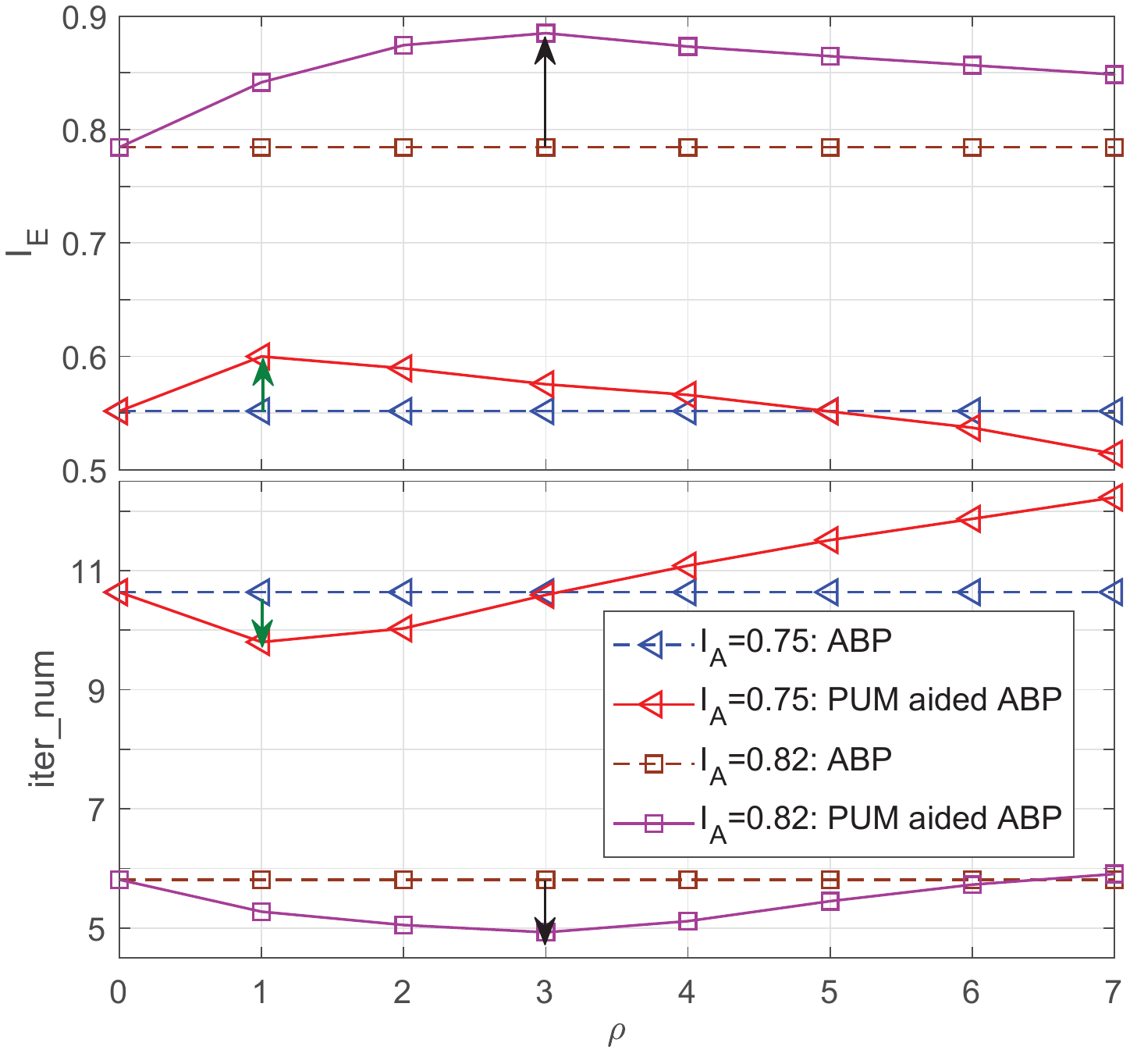}}
     \subfigure[]{
    \label{fig:selected P_Imax} 
    \includegraphics[width=80mm,height=65mm]{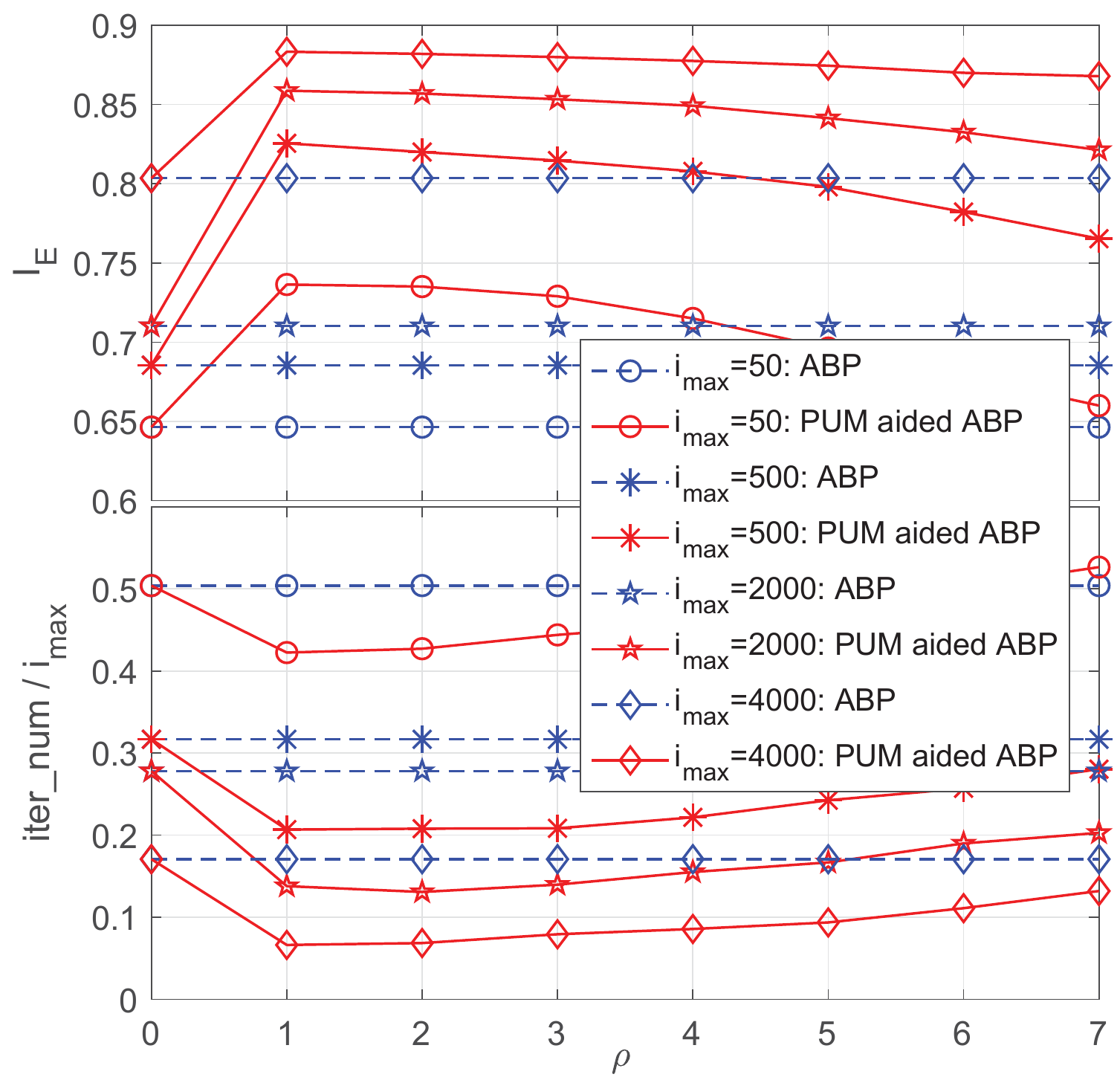}}
   \caption {Extrinsic information analysis for finding good perturbation factor $\rho$ of (127, 92)-BCH code: (a) with $I_A \in \{0.75, 0.82\}$ and $i_{\max}=20$; (b) with $I_A=0.75$ and $i_{\max}\in \{50,500,2000,4000\}$.}
   \label{fig:selected P_BCH}
\end{figure}

Fig. \ref{fig:selected P_BCH} shows an example of extrinsic information analysis for the perturbation factor $\rho$ with (127, 92)-BCH code. Specifically, in Fig. \ref{fig:selected P_IA}, two good values of $\rho$ for $I_A=0.75$ and $I_A=0.82$ are found to be 1 and 3, respectively, both of which give rise to the maximum  $I_E$ and the smallest $iter\_num$. Moreover, the advantages of PUM aided ABP on both the extrinsic information and convergence speed can be observed in Fig. \ref{fig:selected P_Imax} for all $i_{\max}\in \{50,500,2000,4000\}$.

The proposed extrinsic information analysis can be carried out off-line and provides a  simple but effective method to select the perturbation factor for the decoder.

%

\subsection{P-ABP with Partial Layered Scheduling (PL-P-ABP)}
\subsubsection{Layered Scheduling for HDPC Matrix}
As introduced in Section I, for LDPC codes, both the shuffled BP and the layered BP have convergence rates twice faster than the flooding BP by timely 
propagating the latest updated messages. However, due to the special structure of parity-check matrix  ${\bf{ \bar {H}}}_b^{(i)}$ after GE, the flooding, shuffled and layered based ABP exhibit  different performances in HDPC codes, compared to LDPC codes.
\begin{figure}[!t]
  \centering
   \includegraphics[width=88mm,height=90mm]{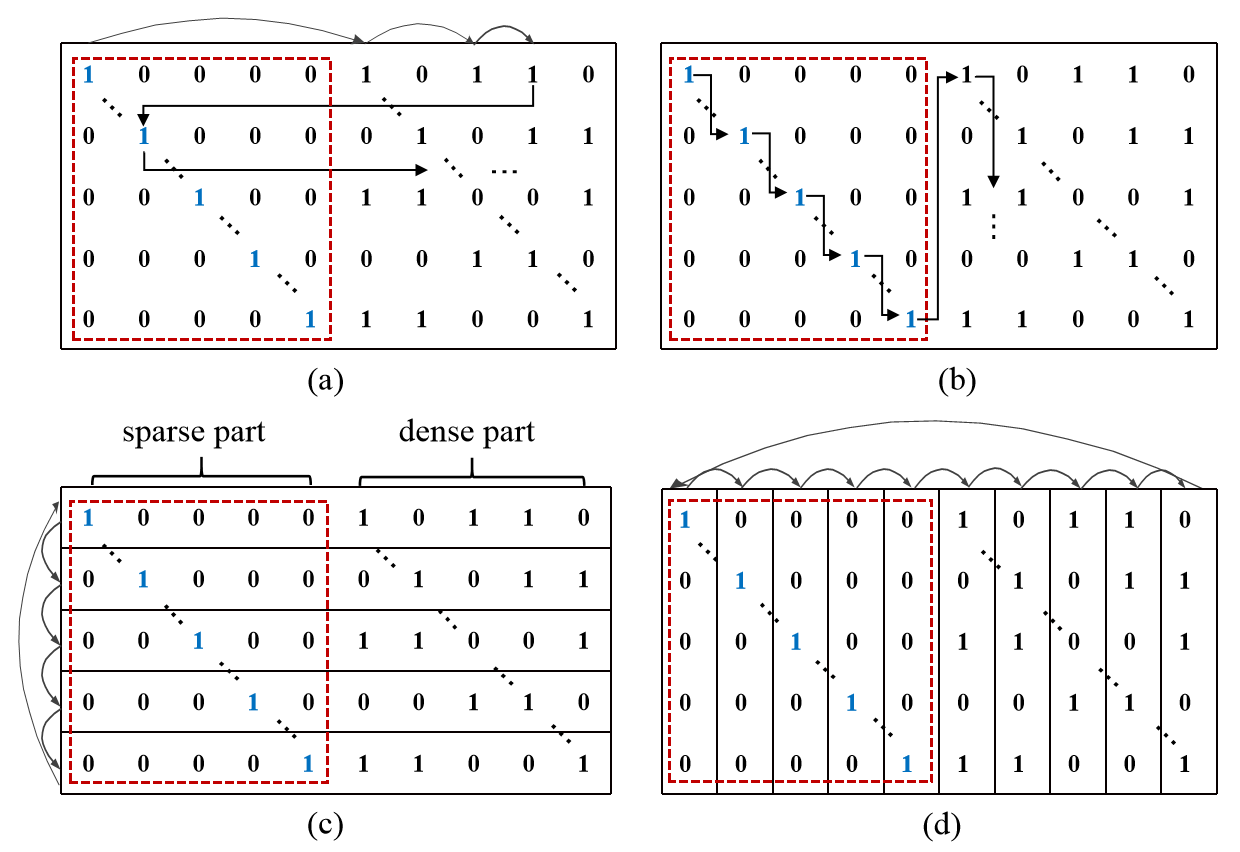}
   \caption {Schematic diagrams of ABP algorithms based on different scheduling strategies: (a) C2V  update of flooding ABP (Phase 1); (b) V2C update of flooding ABP (Phase 2); (c) C2V and V2C updates of layered ABP; (d) C2V and V2C updates of shuffled ABP.
     }\label{fig:three schedules}
\end{figure}
\begin{figure}[!t]
  \centering
   \subfigure[]{
    \label{fig:three schedules compare FER} 
    \includegraphics[width=80mm,height=60mm]{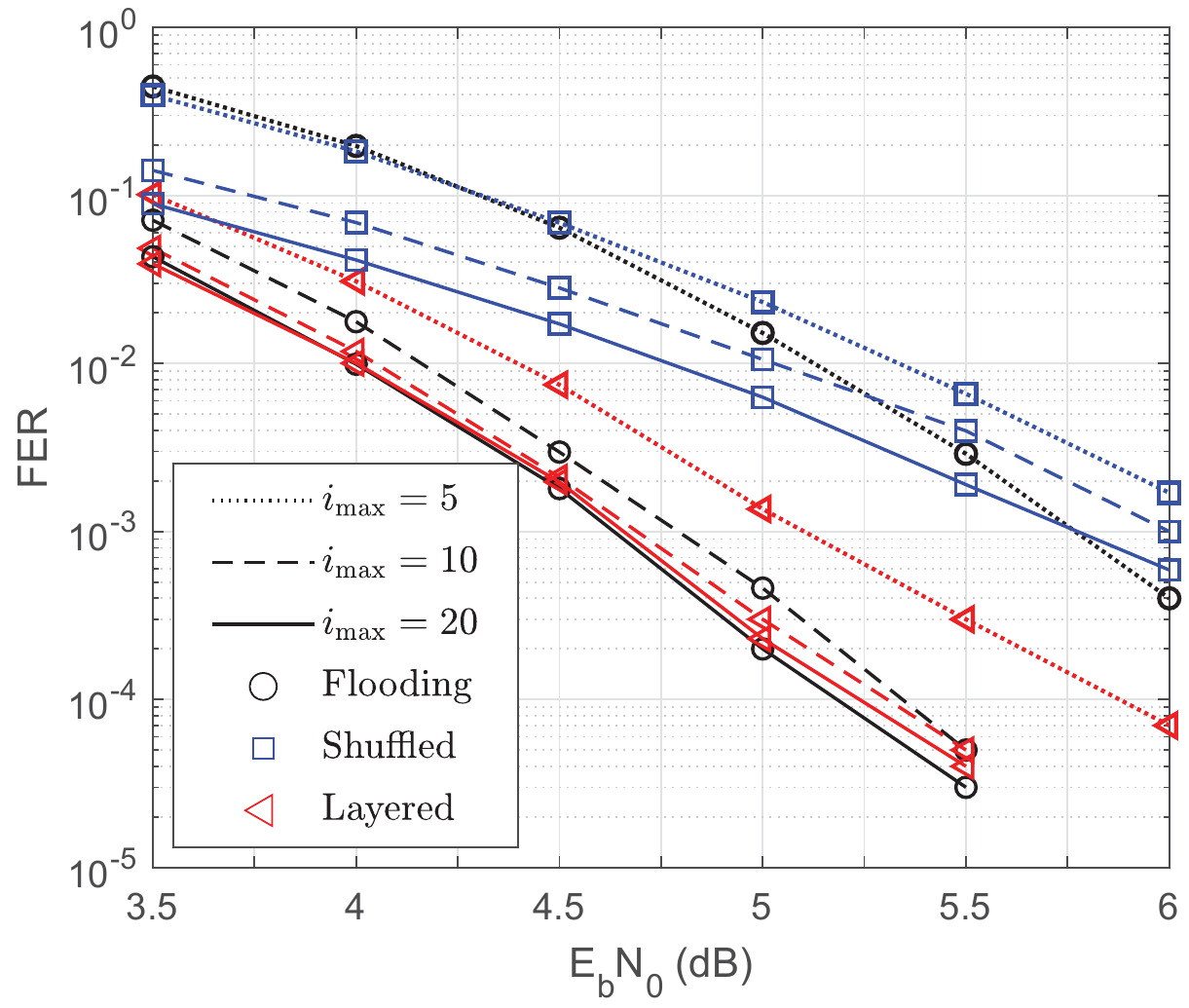}}
     \subfigure[]{
    \label{fig:three schedules compare IER} 
    \includegraphics[width=80mm,height=60mm]{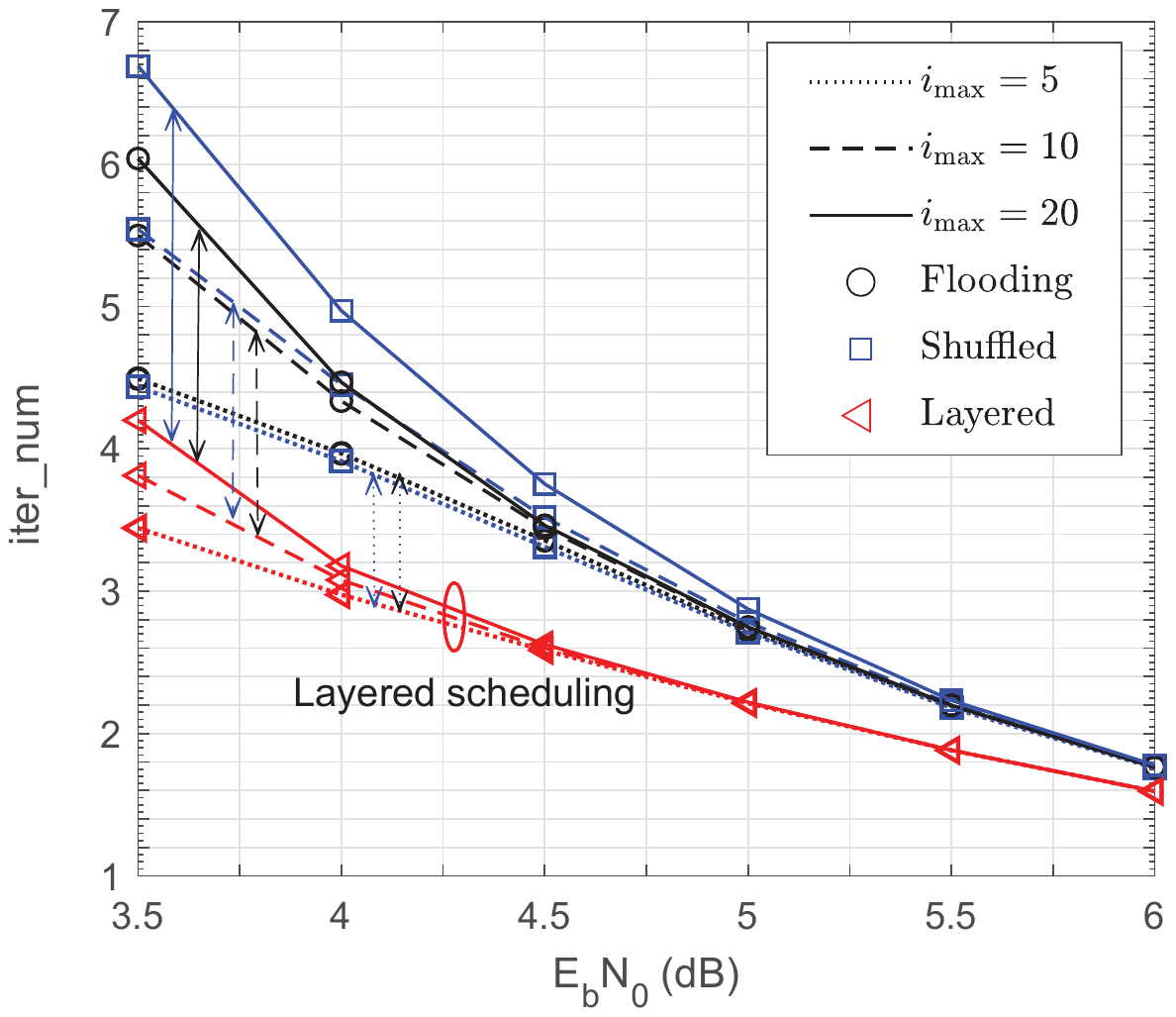}}
   \caption {Performance comparison of different scheduling schemes for ABP with (31, 25)-RS code: (a) FER performance; (b) Average iteration number required to achieve FER shown in Fig. \ref{fig:three schedules compare FER}.}
   \label{fig:three schedules compare}
\end{figure}

Fig. \ref{fig:three schedules} shows the schematic diagrams of the ABP algorithms with different scheduling strategies. In each ABP iteration, the original parity-check matrix  ${\bf{ H}}_b$ of RS or BCH codes is transformed into a systematic matrix, i.e., ${\bf{ \bar {H}}}_b^{(i)}$, where the diagonal sub-matrix (shown within the red dashed rectangle) and the complementary sub-matrix are called the sparse part and the dense part of the transformed matrix, respectively. Fig. \ref{fig:three schedules}(a) and Fig. \ref{fig:three schedules}(b) show the C2V and V2C message updating of flooding ABP, respectively; Fig. \ref{fig:three schedules}(c) and Fig. \ref{fig:three schedules}(d) individually  describe the layered and shuffled ABP. {As seen in Fig. \ref{fig:three schedules}(c), both the C2V and V2C updates of the layered ABP are similar to the row-by-row C2V update of flooding ABP (see arrows shown in Figs. \ref{fig:three schedules}(a) and \ref{fig:three schedules}(c))}. In each row of the layered scheduling iteration, all the C2V and V2C messages of the neighboring VNs can be updated, including the sparse part and the dense part. On the other hand, {both the C2V and V2C updates of the shuffled ABP in Fig. \ref{fig:three schedules}(d) are similar to the column-by-column V2C update of flooding ABP (see arrows shown in Figs. \ref{fig:three schedules}(b) and \ref{fig:three schedules}(d))}. Again, it is stressed that for each VN in the sparse part, there is one neighboring CN only. Due to the column-by-column iteration pattern, there is no V2C  update in the sparse part for both the flooding and shuffled ABP according to (\ref{Eq:flooding_v2c}); and only once C2V update to the single VN in each column of the sparse part for the shuffled ABP according to (\ref{Eq:_shuffled c2v}).

 Fig. \ref{fig:three schedules compare} shows the error correction performances and the convergence rates of different scheduling based ABP algorithms with maximum iteration numbers $i_{\max}$ of 5, 10, and 20, respectively.  
 As seen from both Fig. \ref{fig:three schedules compare FER} and Fig. \ref{fig:three schedules compare IER}, the layered ABP (red lines) can achieve similar or better FER performance with lesser iteration counts, compared to the other two scheduling algorithms. Its superiority is more noticeable in low SNR region as $i_{\max}$ increases. The FER performance of shuffled scheduling is worse than the other two scheduling strategies when $i_{\max}$ is small.

Based on the above analysis, we adopt the layered scheduling as it enjoys the fastest convergence rate and similar or better error correction performance within a certain $i_{\max}$. Accordingly, the extrinsic LLR in (\ref{Eq:LLR_ext}) is rewritten as

\begin{small}
\begin{equation}
\label{Eq:LLR_ext_layered}
  \begin{split}
  L_{\text{ext}}^{(i)}({v_n})\!=\!\sum\limits_{\substack{c_m \in {\cal{M}}\left(v_n \right)\\c_{j}<c_m}}\!{2{{\tanh }^{ - 1}}\left( {\prod\limits_{v_j  \in {\cal{N}}\left( c_m \right)\backslash v_n} {\tanh \left( {\frac{{{L^{(i)}}({v_j })}}{2}} \right)} } \right)}\\
  \!+\!\sum\limits_{\substack{c_m \in {\cal{M}}\left(v_n \right)\\c_{j}>c_m}}\!{2{{\tanh }^{ - 1}}\left( {\prod\limits_{v_j  \in {\cal{N}}\left( c_m \right)\backslash v_n} {\tanh \left( {\frac{{{L^{(i-1)}}({v_j })}}{2}} \right)} } \right)}
  \end{split}.
\end{equation}
\end{small}

\vspace{0.1in}
\subsubsection{Partial updating}
In BP based decoding, a large number of VNs may converge with large LLRs after a few iterations. For LDPC codes, one may adopt a  forced convergence strategy to significantly reduce the decoding complexity with negligible compromise in error correction performance  \cite{Zimmermann04}. For HDPC codes with a large number of short cycles, some edges of short cycles may be skipped by  partial updating, which is helpful to obtain an enhanced error correction performance \cite{Aslam2017}. However, straightforward application of the partial updating strategy of e-Flooding schedule in\cite{Aslam2017} is not optimal for HDPC codes.


In this work, we propose an improved partial updating strategy combined with layered scheduling for HDPC codes to pursue both fast convergence rate and enhanced error correction performance. In our proposed partial updating strategy,  the edge-state vector $\mathcal{E}$ is iteratively updated to determine which edges should be updated, where
\begin{equation}
\label{eqn_check_rel}
\mathcal{E} = \{\varepsilon_{m,n} \ |\ 1 \leq m \leq M,v_n \in \mathcal{N}\left(c_m\right) \},
\end{equation}
and $\varepsilon_{m,n} \in \{0,1\}$. $\varepsilon_{m,n}=1$ indicates that the edge $\vec{e}_{m,n}$ has not converged, hence will be updated in the next iteration, and vice versa. The improved edge-state update criteria can be described as follows: First, all the edges related with unsatisfied syndromes need to be updated. Next, if the related syndrome $s_{c_m}^{(i)}$ is satisfied, but the minimum LLR magnitude $\eta_{c_m}$ of the connected VNs is smaller than a reliability threshold $\mathcal{T}$, only the edges with indices belong to $\widehat{\mathbf{URL}}^{(i)}$ are updated in the next iteration. The reliability threshold $\mathcal{T}$ is a function of the average CN degree $\bar{d}_c$ which is shown in \cite{Aslam2017}. Different from e-Flooding \cite{Aslam2017}, the amplitude comparison is not required\footnote{In fact, this has been done in PUM. } in our solution to decide which VNs need to be updated in the \textit{i}-th iteration. Only VNs in $\widehat{\mathbf{URL}}^{(i)}$ are selected for updating.

The detailed algorithm of the proposed PL-P-ABP is shown in $\textbf{Algorithm 1}$. In each iteration of $\textbf{Algorithm 1}$, PUM and GE are implemented first from Steps 3 to 5; then the edge-state vector $\mathcal{E}$ is updated from Steps 6 to 14; finally the partial layered updating is implemented according to $\mathcal{E}$, followed by LLRs updating and termination judgment.
\begin{algorithm}[!h]
\NoCaptionOfAlgo
 \caption{\textbf{Algorithm 1}: P-ABP with partial layered scheduling (Proposed PL-P-ABP)}\label{Alg1}
 Initialize ${\bf{L}}^{(0)} = \mathbf{L}_\text{ch} $, $i_{\max}$\;
 \For{$ i = 0: i_{{\max}}-1 $}
 { Sorting ${\bf{L}}^{(i)}$, generate  $\mathbf{URL}^{(i)}$ and $\mathbf{RL}^{(i)}$  \;
   Execute  PUM (\textbf{SubAlg-1}), return $\widehat {\mathbf{URL}}^{(i)}$ \;
   Execute  GE on $\mathbf{H}_b$, return $\mathbf{\overline{H}}^{(i)}_b$ \;
   Update the edge-state vetor $\mathcal{E}$:  \ \ \ \ \ \\ \ \  \
   Set $ \mathcal{E} = \{\varepsilon_{m,n}=0|1 \leq m \leq M, v_n \in \mathcal{N}(c_m)\} $ \;
\For {$  \forall \ c_m , 1 \leq m\leq M $ }
{
    \If {$ s_{c_m}^{(i)} \neq 0 $ }
       {\For {$  \forall \ v_n \in \mathcal{N}(c_m) $}
           {$\varepsilon_{m,n}=1$ \;}
       }

    \If {$  s_{c_m}^{(i)} = 0 \ \& \  \eta_{c_m} < \mathcal{T}  $ }
       {\For {$  \forall \  v_n \in \mathcal{N}(c_m)\cap \widehat{\mathbf{URL}}^{(i)} $}
        { $\varepsilon_{m,n}=1$ \; }
       }
}
  Execute partial layered updating: \ \ \ \ \ \\ \ \  \
   \For{$ m = 1:M $}
   {
        \For{$\forall v_n \in \mathcal{N}(c_m)$}
		{
		 \If{$ \varepsilon_{m,n}=1 $}
		 {
		  update $C_{c_m \rightarrow v_n}^{(i)}$ by  (\ref{Eq:flooding_c2v})\;
          update $V_{v_n \rightarrow c_m}^{(i)}$ by  (\ref{Eq:_layered v2c})\;
		 }
		}	
    }
   Update ${\bf{L}}^{(i)}$ by (\ref{Eq:LLR_ext_layered}) and (\ref{graDes})\;
   Hard decision on ${\bf{L}}^{(i)}$ to yield $\mathbf{\hat{c}}^{(i)}$ \;
   \If {$\{ s_{c_m}^{(i)}=0| 1 \leq m \leq M \}$}
   {Terminate decoding\;}
   Proceed to Step $\mathbf{3}$ \;
   }
  Return $\mathbf{\hat{c}}^{(i)}$ \;
End
\end{algorithm}
\vspace{0.1in}
\subsubsection{Variation of PL-P-ABP for Product Codes}
The product code, as a special concatenation of some linear block codes, such as BCH and RS codes, is a widely recognized technique to attain multiplied minimum distance. 
Constructions of a product code can be found in \cite{pyndiah1994,pyndiah1996,pyndiah1998}.
Consider the product code $\mathfrak{P}=\mathfrak{C}^1 \times \mathfrak{C}^2$ with two component codes $\mathfrak{C}^{j}$ with parameters of $(n_j,k_j,d_j), j=1,2$, where $n_j$, $k_j$ and $d_j$ denote the codeword length, the message length, and the minimum distance, respectively. The codeword length of $\mathfrak{P}$ is $n_p=n_1 \times n_2$, the information bit length is $k_p=k_1 \times k_2$, and the minimum distance is $d_p=d_1 \times d_2$.

TAB \cite{Jego2009} is a turbo-oriented ABP decoding for product codes, in which a parity-check matrix adaptation is moved outside of the ABP decoding loop with very few maximum local iteration number (usually 3 to 5). Moreover, the damping factor is not adopted in TAB. The effect of extrinsic LLR is reduced during the \textit{a posteriori} LLR computation process \cite{pyndiah1994}.

In this work, the proposed PL-P-ABP is further modified to adapt to the turbo iterative decoding of product codes. Such a modified decoding scheme, called PL-P-ABP-P, is specified in $\textbf{Algorithm 2}$, where $i_{\text{global}}$ denotes the maximum global iteration (i.e., outer loop iteration) number, and $i_{\text{local}}$  denotes the maximum local iteration number of PL-P-ABP. $NumSucc$ is the counter of successful decoded rows or columns in each half global iteration, which is used for termination decision. Note that PUM and GE are moved outside of local PL-P-ABP decoding loop. Furthermore, considering the small number of local iterations, the LLR updating in global iterations adopts the following rule to prevent error propagation:
\begin{equation}\label{Eq:TPC LLR_proposed}
{
\mathbf{L}^{(i+1)}=
\begin{cases}
\mathbf{L}^{(i)}, &\mbox{if $\sum S_{c_m}^{(i)} =0$ };\\
\mathbf{L}_\text{ch}, &\mbox{otherwise}
\end{cases},
}
\end{equation}
where $S_{c_m}^{(i)}$ is the syndrome of $m$-th CN of component codes. LLRs are updated only if all the checks of component codes in a certain row or  column are satisfied; otherwise, the initial channel information $\mathbf{L}_\text{ch}$ will be used in the global iterations.
\begin{algorithm}[!h]
\NoCaptionOfAlgo
 \caption{\textbf{Algorithm 2}: Variation of PL-P-ABP for product codes (Proposed PL-P-ABP-P)}\label{Alg2}
 Initialize  $n_1$, $n_2$, ${\mathbf{L}}^{(0)} =\mathbf{L}_\text{ch} $,  $NumSucc=0$ \;
 Set $i_{\text{global}}$, $i_{\text{local}}$ \;
 \For{$i = 0: i_{\text{global}}-1$}
 {
   \For{$ row=0:n_1-1$}
     {Sorting $\mathbf{L}^{(i)}(row,:)$, generate  $\mathbf{URL}^{(i)}$ and $\mathbf{RL}^{(i)}$ \;
     Execute PUM (\textbf{SubAlg-1}), return $\widehat{\mathbf{URL}}^{(i)}$\;
      Execute GE on $\mathbf{H}_b$, return $\overline{\mathbf{H}}_b^{(i)}$ \;
      \For{$j=0:i_{\text{local}}-1$}
       { Execute \textbf{Algorithm 1}\;}
       Hard decision to yield $\mathbf{\hat{c}}^{(i)}(row,:)$ \;
      \If {$\sum S_{c_m}^{(i)} =0$}
      {$NumSucc=NumSucc+1$ \;}
      Update LLR: $\mathbf{L}^{(i+1)}(row,:)$ by (\ref{Eq:TPC LLR_proposed}) \;
     }
      \If { $NumSucc=n_1$}
      {Terminate decoding, proceed to Step $\mathbf{29}$;}
     $NumSucc=0$\;

   \For{$ col=0:n_2-1$}
     {Sorting $\mathbf{L}^{(i)}(:, col)$, generate  $\mathbf{URL}^{(i)}$ and $\mathbf{RL}^{(i)}$ \;
     Execute PUM (\textbf{SubAlg-1}), return $\widehat{\mathbf{URL}}^{(i)}$\;
      Execute GE on $\mathbf{H}_b$, return $\overline{\mathbf{H}}_b^{(i)}$ \;
      \For{$j=0:i_{\text{local}}-1$}
       { Execute \textbf{Algorithm 1}\;}
       Hard decision to yield $\mathbf{\hat{c}}^{(i)}(:,col)$ \;
       \If {$\sum S_{c_m}^{(i)} =0$}
      {$NumSucc=NumSucc+1$ \;}
     Update LLR: $\mathbf{L}^{(i+1)}(:, col)$ by (\ref{Eq:TPC LLR_proposed}) \;
     }
    \If { $NumSucc=n_2$}
      {Terminate decoding, proceed to Step $\mathbf{29}$\;}
      $NumSucc=0$\;
     Proceed to Step $\mathbf{4}$ \;
   }
  Return $\mathbf{\hat{c}}^{(i)}$ \;
End
\end{algorithm}
\begin{algorithm}[!h]
\NoCaptionOfAlgo
 \caption{\textbf{Algorithm 3}: P-ABP with hybrid dynamic scheduling (Proposed HD-P-ABP)}\label{Alg3}
 Initialize ${\bf{L}}^{(0)}=\mathbf{L}_\text{ch}$, $C2V_\text{update}=0$, $i_{\max}$, $total\_edges$\;
 \For{$i = 0: i_{{\max}}-1$}
 { Sorting ${\bf{L}}^{(i)}$, generate  $\mathbf{URL}^{(i)}$ and $\mathbf{RL}^{(i)}$  \;
   Execute PUM (\textbf{SubAlg-1}), return $\widehat {\mathbf{URL}}^{(i)}$ \;
   Execute GE on $\mathbf{H}_b$, return $\mathbf{\overline{H}}^{(i)}_b$ \;
   Execute layered scheduling once for bits in $\widehat {\mathbf{URL}}^{(i)}$: \ \ \ \ \ \ \ \ \ \ \ \ \ \ \ \ \

    \For {$m=1:M$}
  {
  Generate and propagate $C_{c_m \rightarrow v_n}^{(i)}$ with (\ref{Eq:flooding_c2v});
  }

  Execute D-SVNF scheduling: \ \ \ \ \ \ \ \ \ \ \ \ \ \ \ \ \ \ \ \ \ \ \
 \While{$\left( C2V_\text{update} < total\_edges-M \right)$}
 {

   \If{all $\{u_n=1, 1 \leq n \leq N\}$    \label{node_selection}}
	{ Reset $\{u_n = 0$, $1 \leq n \leq N\}$ \;}

   Select $v_p:u_p=0, u_i=1, \forall i<p$. Set $u_p = 1$ \;
     \For{$c_m \in \mathcal{M}\left(v_p\right), s_{c_m}=1, v_n \in \mathcal{N}\left(c_m\right)\setminus v_p$ \label{res_compute}}
	  {
        Compute $R^{(i)} \left( {c_m \rightarrow v_n} \right)$ with (\ref{Eq:residual})\;}		

	  Find $R^{(i)} \left( {c_i \rightarrow v_j} \right) = {\max} \{ R^{(i)} \left({c_m \rightarrow v_n}\right) |$\\ $c_m \in \mathcal{M}\left(v_p\right), v_n \in \mathcal{N}\left(c_m\right)\setminus v_p \}$ \;

   Generate and propagate $C_{c_i \rightarrow v_j}^{(i)}$ with (\ref{Eq:flooding_c2v}) \;
   $C2V_\text{update}=C2V_\text{update}+1$ \;
   Set $R^{(i)} \left( c_i \rightarrow v_j \right) = 0 $\;

 \For{$c_a \in \mathcal{M}\left(v_j\right)\setminus c_i$}
   {
	 Generate and propagate $V_{v_j \rightarrow c_a}^{(i)}$ with (\ref{Eq:_layered v2c})\;	
   }
   \eIf{$u_j=1$}
	 {Proceed to Step~\ref{node_selection}  \;}
	 {Set $v_p=v_j$, $u_p=1$. Proceed to Step~\ref{res_compute}\;}

 }
   Update ${\bf{L}}^{(i)}$ by (\ref{Eq:LLR_ext_layered}) and (\ref{graDes})\;
   Hard decision on ${\bf{L}}^{(i)}$ to yield $\mathbf{\hat{c}}^{(i)}$ \;
   \If {$\{ s_{c_m}^{(i)}=0| 1 \leq m \leq M \}$}
   {Terminate decoding\;}
   Proceed to Step $\mathbf{3}$ \;
   }
  Return $\mathbf{\hat{c}}^{(i)}$ \;
End
\end{algorithm}

\begin{table*}[!t]
\newcommand{\tabincell}[2]{\begin{tabular}{@{}#1@{}}#2\end{tabular}}
\centering
\begin{threeparttable}
\renewcommand{\arraystretch}{1.15}
\caption{Complexity analysis of fixed and dynamic scheduling based algorithms (per iteration)}
\label{table_1}

\begin{tabular}{|c|c|c||c|c|c|c|}

\hline

\bfseries Scheme & \bfseries Code  & \bfseries Algorithm & \bfseries V2C update & \bfseries C2V update & \bfseries Residual computation & \bfseries Real-value comparison \\ \hline
\multirow{5}{*}{fixed} & \multirow{3}{*}{\tabincell{c}{RS/BCH\\codes}} & ABP \cite{Jiang2006}  & $E$ & $E$ & $0$ & $0$\\ \cline{3-7}

    &  & e-Flooding ABP \cite{Aslam2017} & $\leq E$ & $\leq E$ & $0$ & $\psi$\\ \cline{3-7}

    &  & PL-P-ABP (Proposed) & $\leq E$ & $\leq E$ & $0$ & $\leq N+K(K-1)/2$\\ \cline{2-7}

   & \multirow{2}{*}{\tabincell{c}{product\\codes}}  & TAB \cite{Jego2009} & $n_1E_2+n_2E_1$ & $n_1E_2+n_2E_1$ & $0$ & $0$\\ \cline{3-7}

    &  & PL-P-ABP-P (Proposed) & $\leq (n_1E_2+n_2E_1)$ & $\leq (n_1E_2+n_2E_1) $ & $0$ & \tabincell{c}{$\leq n_1(n_2+k_2(k_2-1)/2)$\\$+n_2(n_1+k_1(k_1-1)/2)$}\\\hline\hline

\multirow{2}{*}{dynamic}  &  \multirow{2}{*}{\tabincell{c}{RS/BCH\\codes}} & LRBP \cite{Lee2015}  & $\left(\bar{d_v}-1\right)E$ & $E$ & $\left(\bar{d_v}-1\right)\left(\bar{d_c}-1\right)E$ & $\left(\bar{d_c}-1\right)E$\\ \cline{3-7}

& & {DP-RBP \cite{Lee2017}}  & {$\left(\bar{d_v}-1\right)E$} & {$E$} & {$\left(\bar{d_v}-1\right)\left(\bar{d_c}-1\right)E$} & {$\left(\bar{d_c}-1\right)E$}\\

\cline{3-7}

& & HD-P-ABP  (Proposed) & $\left(\bar{d_v}-1\right)E$ & $E$ & $\leq \left(\bar{d_v}-1\right)\left(\bar{d_c}-1\right)E$ & $\leq \left[ \bar{d_v} \left(\bar{d_c}-1\right) -1 \right]E$\\\hline

\end{tabular}
\begin{tablenotes}
      \footnotesize
      \item  \text{ }\text{ } \text{ }\text{ } \text{ }\text{ }\text{ }  \text{ } \text{ }$E$: total edges in the TG \text{ }\text{ } \text{ }\text{ }\text{ }\text{ } \text{ } \text{ }$\bar{d_v}$: average VN degree   \text{ }\text{ } \text{ }\text{ }\text{ }\text{ }\text{ }\text{ }$\bar{d_c}$: average CN degree  \text{ } \text{ } \text{ }\text{ }\text{ }\text{ }\text{ }\text{ }$\psi$: see \cite{Aslam2017}

      \item \text{ }\text{ }\text{ }\text{ }\text{ }\text{ }\text{ }\text{ }\text{ }\text{ }\text{ }\text{ } $n_1, n_2, k_1, k_2, E_1, E_2$: codeword lengths, message lengths, and  total edges in the TG of product component codes
    \end{tablenotes}
\end{threeparttable}
\end{table*}

\vspace{0.1in}

\subsection{P-ABP with Hybrid Dynamic Scheduling (HD-P-ABP)}
 Partial layered scheduling is a form of fixed scheduling, which means that the scheduling order/sequence can be predetermined off-line. Compared with fixed scheduling, dynamic scheduling can further improve the convergence rate, by determining the scheduling order on-the-fly during decoding runtime.  The main idea of dynamic scheduling is to first update the message associated with the largest LLR residual (then the second largest and so on). Formal definition of ``LLR residual" can be found in (\ref{Eq:residual}). This will allow the decoder to focus on the part of TG that has not converged. Moreover, dynamic decoding has potential to circumvent trapping sets in TG and hence improve the error correction performance \cite{Casado2010}.

 However, dynamic scheduling may lead to certain types of greedy groups, each consisting of a few CNs or VNs which excessively consuming the computing resources. A greedy group is called a myopic error if it is formed by a small number of CNs \cite{Casado2010}.  The greedy group formed by certain VNs may also result in some silent-variable-nodes which never have a chance to be updated in the decoding process \cite{SVNF2013}.  Although some improved dynamic schemes have been proposed to prevent those greedy groups for LDPC codes \cite{SVNF2013,DSVNF2017,Casado2010}, little is understood on how to prevent greedy groups for HDPC codes. The LRBP for RS codes in \cite{Lee2015} can efficiently avoid myopic error by sequentially updating the C2V message, but may not be able to prevent the silent-variable-nodes. Motivated by these issues, we propose a hybrid dynamic scheduling for HDPC codes to circumvent both types of greedy groups. 

 Our proposed HD-P-ABP is a hybrid schedule that combines layered scheduling and D-SVNF scheme. The main idea of HD-P-ABP is that the layered scheduling is applied only once for the unreliable bits in $\widehat{\mathbf{URL}}^{(i)}$ before the D-SVNF scheduling. By doing so, one can prevent both the myopic error by layered  scheduling, and the silent-variable-nodes by D-SVNF scheduling. Moreover, it provides one more chance of C2V message updating for the unreliable bits by layered scheduling. Those updated messages before the second iteration are almost independent and reliable for a HDPC code with large number of short cycles \cite{Mao2001}.

The details of HD-P-ABP are presented in \textbf{Algorithm 3} with some definitions given first. The C2V message residual $R^{(i)} \left( {c_m \rightarrow v_n} \right)$ is defined as
\begin{equation}
\label{Eq:residual}
R^{(i)} \left( {c_m \rightarrow v_n} \right) = \left | C_{c_m \rightarrow v_n}^{(i-1)} - C_{c_m \rightarrow v_n}^{(i)} \right |,
\end{equation}
where $C_{c_m \rightarrow v_n}^{(i-1)}$ denotes the C2V message at the $(i-1)$-th iteration. To prevent the silent-variable-nodes, a binary vector $\mathbf{u}=\{u_n, 1 \leq n \leq N \}$ is set to record the update states of VNs, where $u_n=1$ indicates that the variable node $v_n$ has been updated and hence should not be updated again in the current iteration. The variable $total\_edges$ represents the total edges in the TG, and
$C2V_{\text{update}}$ is used to count the C2V update numbers, which  should not be larger than $total\_edges$.

In \textbf{Algorithm 3}, PUM and GE are implemented first. From Steps 6 and 7, the C2V messages of unreliable bits in $\widehat{\mathbf{URL}}^{(i)}$ are updated once with layered scheduling. By doing so, all the CNs have been updated at least once to prevent the myopic error. Then D-SVNF scheduling is implemented from Steps 8 to 24.
It is noted that D-SVNF allows a dynamic updating order of VNs, which can not only get rid of the silent-variable-nodes, but also prioritize the message updating of the most probably erroneous VNs which are connected with those CNs with unsatisfied syndromes, thus resulting in a faster decoding convergence rate \cite{DSVNF2017}.

\section{Complexity Analysis}
In this section, the decoding complexities of the proposed and prior-art algorithms per decoding iteration are analyzed. Table \ref{table_1} lists the total numbers of V2C and C2V message updates, residual computations, and real-value comparisons in a single decoding iteration. For product codes, ``per iteration'' means per global iteration.  The real-value comparison is used to measure the scheduling complexity which can be realized in the hardware by a full-adder circuit \cite{Damle2013}.
For fair comparison, no HDD is considered to assist SDD in all the compared algorithms in Table \ref{table_1}.


For fixed algorithms, ABP \cite{Jiang2006} needs $E$ numbers of V2C and C2V message updates, where $E$ denotes the number of total edges in TG. Both e-Flooding ABP \cite{Aslam2017} and the proposed PL-P-ABP require fewer numbers of V2C/C2V updates than ABP, owing to partial updating strategies. The real-value comparisons of e-Flooding ABP denoted as $\psi$ are used for identification of \textit{minimum-reliability} VNs. The approximation of $\psi$ is specified in \cite{Aslam2017}. Compared with $\psi$, the real-value comparisons of proposed PL-P-ABP are quite simple, only involving $N$ numbers of sign comparison and $g(g-1)/2$ numbers of magnitude comparison in the PUM, where $g$ is the number of alternating-sign bits ($g\leq K$). As for product codes, since the component codes are iteratively decoded by local algorithms (i.e., local ABP for TAB, and local PL-P-ABP for PL-P-ABP-P), the overall V2C/C2V updates and real-value comparisons of TAB \cite{Jego2009} and PL-P-ABP-P are linear combinations of those of local algorithms. Thus, the V2C and C2V updates of PL-P-ABP-P are less complex than TAB.

{For dynamic algorithms, the proposed HD-P-ABP  has the same V2C and C2V message update counts with LRBP \cite{Lee2015} and {DP-RBP \cite{Lee2017}}, i.e., $\left(\bar{d_v}-1\right)E$ and $E$, respectively, where $\bar{d_v}$ denotes the average VN degree. Since LRBP and {DP-RBP} update CNs with a sequential order and a double-polling mechanism, respectively, which have fixed numbers of residual computation and real-value comparison, i.e., $\left(\bar{d_v}-1\right)\left(\bar{d_c}-1\right)E$ and $\left(\bar{d_c}-1\right)E$, respectively. However, 
the updating order of VNs in the proposed HD-P-ABP is dynamic, in that the residual computation and real-value comparison is implemented on-demand only for the CNs with unsatisfied syndromes (which are connected to the current updating VNs). Therefore, the proposed HD-P-ABP has less residual computational complexity than LRBP and {DP-RBP}. The real-value comparison of HD-P-ABP is less than $\left[\bar{d_v}\left(\bar{d_c}-1\right)-1\right]E$, which is comparable to that of LRBP and DP-RBP.}
\begin{figure*}[!t]
  \centering
   \subfigure[]{
    \label{fig:RS FER performance}
    \includegraphics[width=85mm,height=90mm]{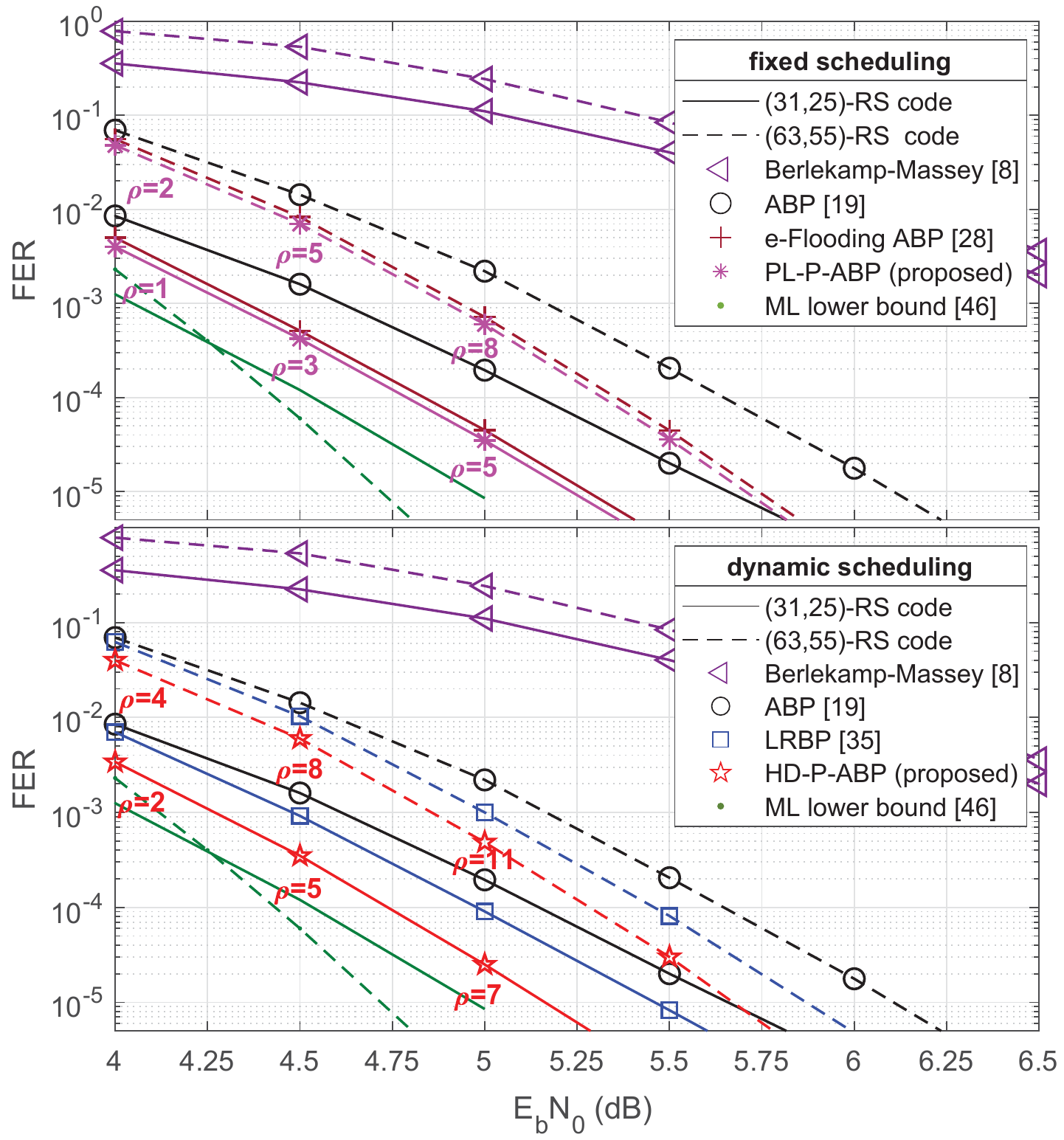}}
     \subfigure[]{
  \label{fig:RS IER performance}
    \includegraphics[width=85mm,height=90mm]{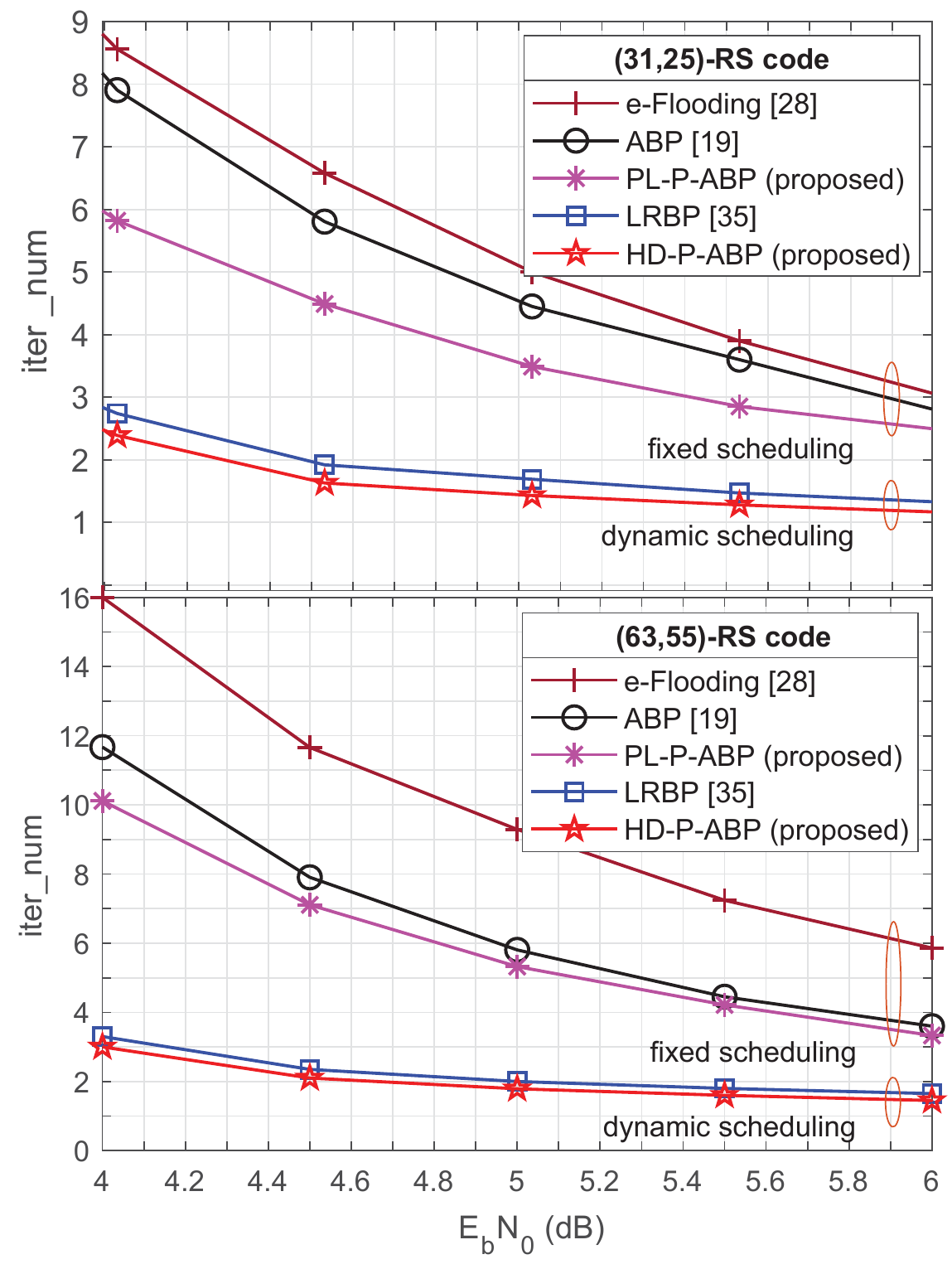}}
   \caption {{FER and decoding convergence comparisons for (31, 25)-RS code  and  (63, 55)-RS code with $i_{\max}=50$: (a) FER performance; (b) Average iteration number required to achieve FER shown in Fig. \ref{fig:RS FER performance}.}}
   \label{fig:RS performance}
\end{figure*}
\section{Simulation results}
In this section, we benchmark the performances of the proposed PL-P-ABP and HD-P-ABP algorithms for RS and BCH codes against the prior-art ABP algorithms, including traditional ABP in \cite{Jiang2006}, LRBP in \cite{Lee2015}, {DP-RBP in \cite{Lee2017}}, and e-Flooding ABP in \cite{Aslam2017}. For product codes, the proposed PL-P-ABP-P is compared with the Chase-Pyndiah algorithm in \cite{pyndiah1998}, and TAB algorithm in \cite{Jego2009}.  
\subsection{Performance Comparison for RS Codes}
Two types of RS codes are considered for simulation, i.e., rate-0.80 (31, 25)-RS code and rate-0.87 (63,55)-RS code, both having high coding rates, short block lengths, and HDPC matrices with large numbers of short cycles. {The maximum decoding iteration number is set to be $i_{\max}=50$, which follows the setting of \cite{Aslam2017}}.

Fig. \ref{fig:RS performance} compares the  FER performances and the average required iteration numbers ($iter\_num$) of different algorithms for RS codes. The green lines in Fig. \ref{fig:RS FER performance} represent the lower bounds of the maximum likelihood (ML) decoding from \cite{ML1999} and \cite{Jiang2006}, 
the purple lines show the  performance of Berlekamp-Massey decoding algorithm \cite{Massey1969}. For the fixed schemes, e-Flooding ABP achieves better error correction performance with a certain loss of convergence rate compared with ABP (due to the partial updating strategy). The proposed PL-P-ABP can further improve the average convergence rates of ABP by 22.76\% for (31,25)-RS code and 18.5\% for (63,55)-RS code, respectively, while maintaining an error rate performance comparable to that of e-Flooding ABP but about 0.3 dB gains over ABP at FER of $10^{-5}$. Both the two dynamic schemes have comparable rapid decoding convergence rates. The proposed HD-P-ABP can improve the convergence rate of ABP by 67\%, while obtaining enhanced  error rate performance about 0.5 dB and 0.3 dB gains over ABP and LRBP at FER of $10^{-5}$ for (31,25)-RS code, respectively.

\begin{figure}[!t]
  \centering
  \includegraphics[width=85mm,height=85mm]{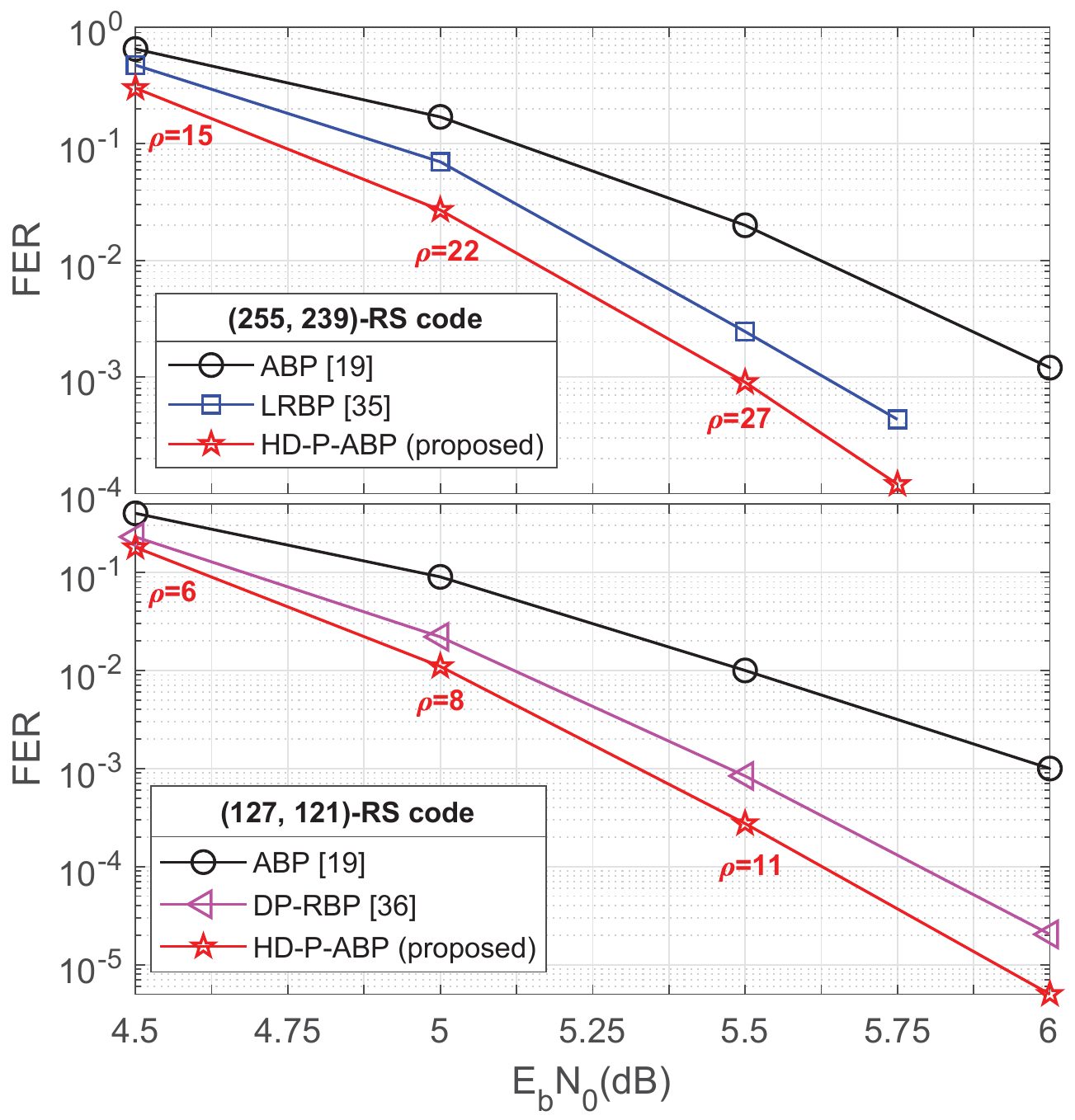}
   \caption {{FER comparison for (255, 239)-RS code and (127, 121)-RS code with $i_{\max}=60$ and algebraic HDD assistance.}}\label{fig:RS FER comp with DP_RBP}
\end{figure}

{
We further bench mark the proposed dynamic scheme HD-P-ABP with LRBP \cite{Lee2015} and DP-RBP \cite{Lee2017} for (255,239)-RS code and (127,121)-RS code, which is shown in Fig. \ref{fig:RS FER comp with DP_RBP}. 
Considering the long codeword length and high decoding complexity of (255,239) and (127,121) RS codes,  the algebraic HDD is also incorporated into the proposed HD-P-ABP for faster decoding and enhanced error rate performances \cite{Lee2015,Lee2017}. As shown in Fig. \ref{fig:RS FER comp with DP_RBP}, with a maximum iteration number of $i_{\max}=60$, the proposed HD-P-ABP  can achieve about 0.5 dB and 0.1 dB of SNR gain over ABP and LRBP for (255,239)-RS code at FER=$10^{-3}$, respectively, and about 0.65 dB and 0.15 dB of SNR gain over ABP and DP-RBP for (127,121)-RS code at FER=$10^{-3}$, respectively. There are two main factors which contribute to the performance gains of the proposed HD-P-ABP, i.e., the proposed PUM strategy and the hybrid dynamic scheduling.
}

\subsection{Performance Comparison for BCH Codes}

\begin{figure*}[!t]
  \centering
   \subfigure[]{
    \label{fig:BCH FER performance}
    \includegraphics[width=80mm,height=85mm]{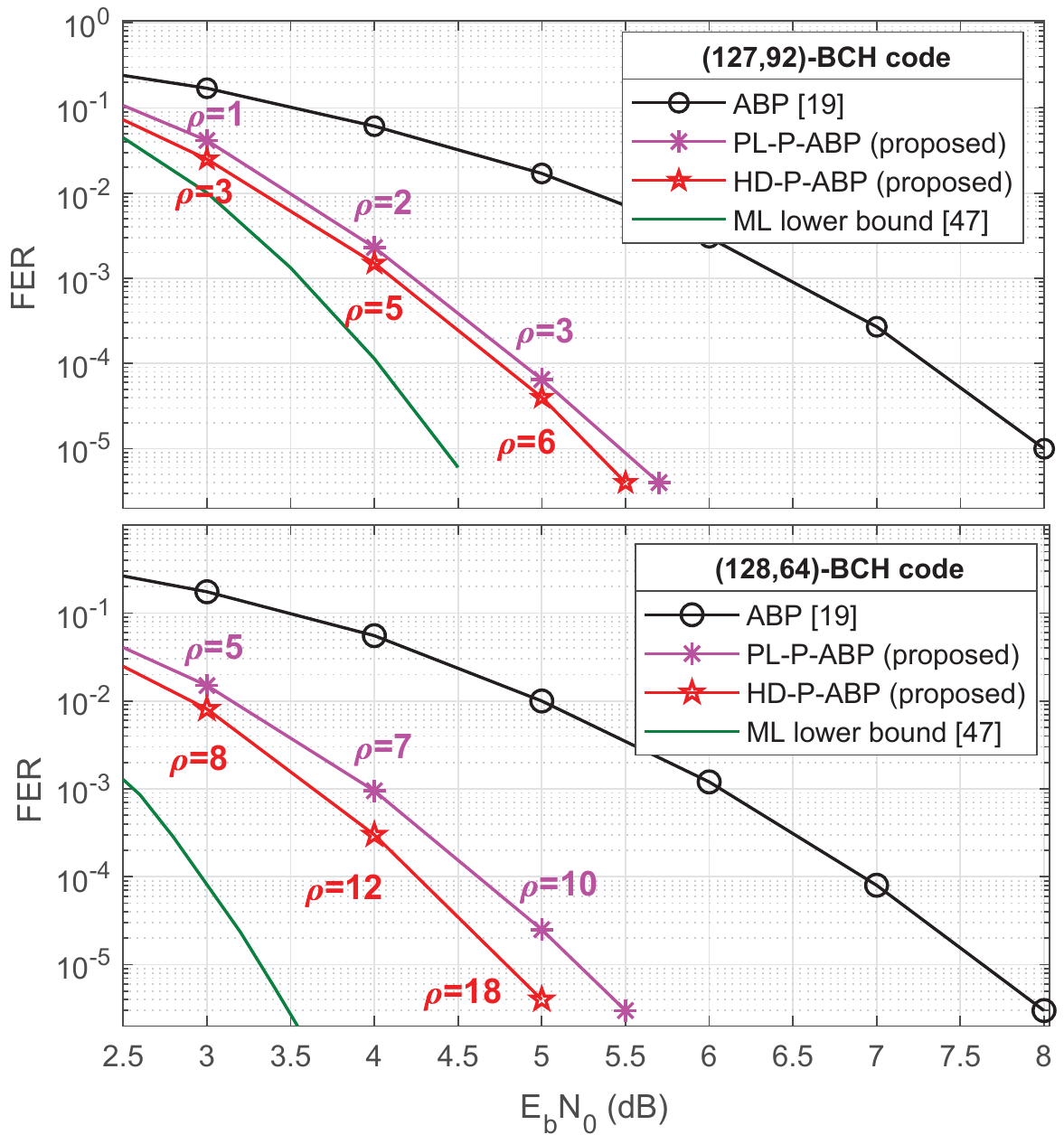}}
     \subfigure[]{
  \label{fig:BCH IER performance}
    \includegraphics[width=80mm,height=85mm]{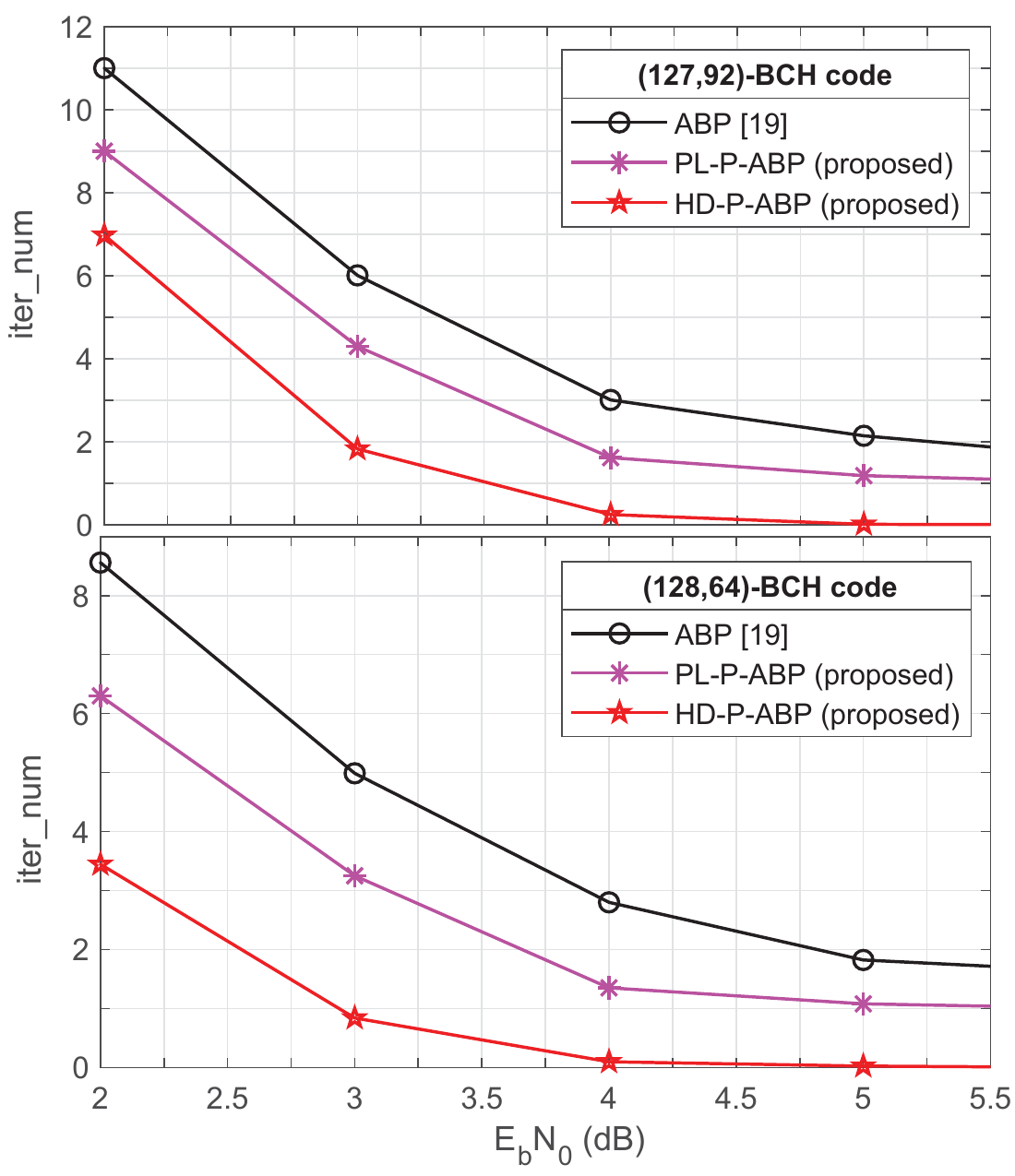}}
   \caption {FER and decoding convergence comparisons for  (127, 92)-BCH code and (128, 64)-BCH code with $i_{\max}=20$: (a) FER performance; (b) Average iteration number required to achieve FER shown in Fig. \ref{fig:BCH FER performance}.\label{fig:BCH performance}}
   \label{fig:BCH performance}
\end{figure*}

In this section, we apply the proposed algorithms to rate-0.72 (127, 92)-BCH code and rate-0.5 (128, 64)-BCH code, and compare them with the traditional ABP in \cite{Jiang2006}. {The maximum iteration number is set to be $i_{\max}=20$ for FER convergence of the proposed algorithms}.

Fig. \ref{fig:BCH performance} shows the FER and convergence rate comparisons for BCH codes. The green lines represent the ML simulation results from \cite{ML2018}. For (127, 92)-BCH code, PL-P-ABP and HD-P-ABP lead to 34.43\% and 74.50\% faster convergence rates compared with ABP, respectively; while for (128, 64)-BCH code, the average convergence rate improvements are severally 38.38\% and 84.49\%. Owing to their faster convergence rates, the proposed algorithms can achieve at least 2.25 dB gains over ABP at FER of $10^{-5}$ within a small iteration number. HD-P-ABP outperforms PL-P-ABP about 0.2-0.4 dB.

\subsection{Performance Comparison for Product Codes}
\begin{figure}[!t]
  \centering
  \includegraphics[width=80mm,height=70mm]{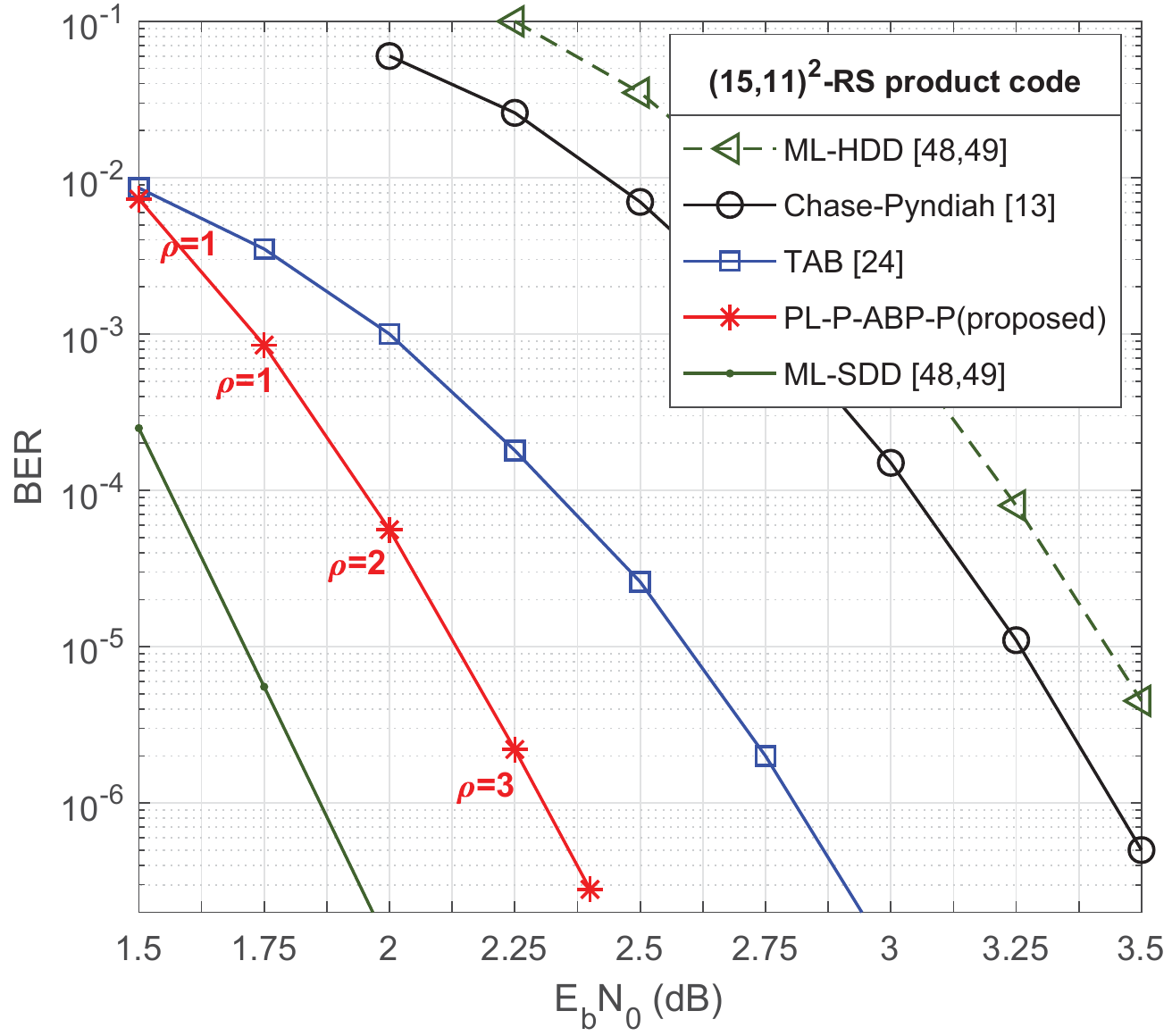}
   \caption {{BER comparison of rate-0.54 $(15, 11)^2$-RS product code  with different decoding methods, where $i_{\text{global}}=8$, $i_{\text{local}}=5$ are set in \cite{Jego2009} and the proposed PL-P-ABP-P.}  \label{fig: TPC BER performance}}
\end{figure}

We consider the rate-0.54 $(15,11)^2$-RS product code. The iteration numbers are set the same as that in \cite{Jego2009}, i.e., global iterations of $i_{\text{global}}=8$ and local iterations of $i_{\text{local}}=5$. Fig. \ref{fig: TPC BER performance} shows the bit error rates (BER) of compared algorithms. {The green solid and dashed lines represent the Poltyrev tight bounds on ML soft decision decoding (ML-SDD) and ML hard decision decoding (ML-HDD), respectively, from \cite{Elkhamy2005} and \cite{Elkhamy2006}.} The proposed PL-P-ABP-P achieves about 0.5 dB and 1.1 dB gains over TAB \cite{Jego2009} and Chase-Pyndiah algorithm \cite{pyndiah1998} at BER of $10^{-6}$, respectively. Moreover, the proposed PL-P-ABP-P is less than 0.5 dB {from the tight bound of ML-SDD} at BER of $10^{-6}$, which is a good result not attainable before. On the other hand, as shown in Table \ref{table_1}, the proposed PL-P-ABP-P also has lower decoding complexity than TAB. Note that the local PL-P-ABP is repeatedly implemented in the global iterations of PL-P-ABP-P, leading to  faster convergence rate than local ABP based TAB (as shown in Fig. \ref{fig:RS IER performance}). Thus, the superiority of PL-P-ABP-P on the decoding convergence rate over TAB follows.
\section{Conclusion}
This paper {considers} the SISO decoding of HDPC codes such as BCH codes, RS codes, and product codes using {an} adaptive belief propagation (ABP) algorithm. 
We have observed that the traditional ABP, which sparsifies the parity-check columns corresponding to the $M$ least reliable bits {(where $M$ denotes the number of parity-check bits), may not be optimal}. Our key idea, called Perturbed ABP (P-ABP), is to include a few unstable bits with large LLR magnitudes but incorrect signs in the parity-check matrix sparsification. We have proposed an off-line extrinsic information analysis method to search the proper number and the locations of these unstable bits.

We then augment the P-ABP with two novel scheduling schemes, namely, partial layered scheduling or hybrid dynamic scheduling resulting in PL-P-ABP (\textbf{Algorithm 1}) and HD-P-ABP (\textbf{Algorithm 3}), respectively. {PL-P-ABP adopts layered scheduling and an improved partial updating strategy to skip unreliable message passing caused by some short cycles in HDPC codes, while HD-P-ABP adopts a hybrid usage of layered scheduling and D-SVNF to circumvent multiple types of greedy groups. To decode product code, we have extended PL-P-ABP to PL-P-ABP-P (\textbf{Algorithm 2}) in a non-trivial way to avoid error propagation between the horizontal and vertical decoding processes.}

Our simulations have shown that PL-P-ABP and HD-P-ABP achieve {gains of 2.25 dB and 2.75 dB, and convergence rate improvement of 38.38 \% and 84.49\% over the traditional ABP for the (128,64)-BCH code, respectively.  Separately, the decoding performance of the proposed PL-P-ABP-P on $(15,11)^2$-RS product code is within 0.5 dB from the tight bound of ML decoding, which is much improved compared to Chase-Pyndiah decoding or TAB decoding.}

{The excellent decoding performance of proposed P-ABP algorithms do not come with higher computational complexity. As shown in Table \ref{table_1}, the proposed algorithms  have less or similar decoding complexities per iteration compared to the prior-art fixed or dynamic ABP algorithms. Coupled with the fewer iteration numbers required to achieve a particular FER (shown in Fig. \ref{fig:RS performance} and \ref{fig:BCH performance}), the proposed P-ABP decoders can be concluded to have lower overall computational complexity.} 

\ifCLASSOPTIONcaptionsoff
  \newpage
\fi
\bibliographystyle{IEEEtran}
\bibliography{bib}

\end{document}